 \documentclass[traditabstract]{aa}
\usepackage{txfonts}
\usepackage{natbib}
\usepackage[pdftex]{graphicx}
\usepackage{url}
\usepackage[draft]{hyperref}

\def\aj{AJ}%
\def\apj{ApJ}%
\def\apjl{ApJ}%
\def\apjs{ApJS}%
\def\aap{A\&A}%
\def\mnras{MNRAS}%
\def\pasp{PASP}%
\def\nat{Nature}%
\def\s4g{S$^4$G}

\begin{document}

\title{The Occulting Galaxy Pair UGC 3995}
  
\subtitle{Dust properties from HST and CALIFA data.}

\author{
B.W. Holwerda\inst{1}  
\and W.C. Keel\inst{2}    }

\institute{European Space Agency Research Fellow (ESTEC), Keplerlaan 1, 2200 AG Noordwijk, The Netherlands
\and Department of Physics and Astronomy, University of Alabama, Box 870324, Tuscaloosa, AL 35487, USA}
\offprints{B.W. Holwerda, \email{benne.holwerda@esa.int}}

\date{Received 12-3-2013  / Accepted 23-5-2013}

\abstract{

UGC 3995 is an interacting and occulting galaxy pair. UGC 3995B is a foreground face-on spiral and UGC 3995A a bright background spiral with an AGN.
We present analysis of the dust in the disc of UGC 3995B based on archival Hubble Space Telescope WFPC2 and PPAK IFU data from the CALIFA survey's first data release.

From the HST F606W image, we construct an extinction map by modeling the isophotes of the background galaxy UGC 3995A and the resulting transmission through UGC 3995B. 
This extinction map of UGC 3995B shows several distinct spiral extinction features. 
The radial distribution of $A_V$ values declines slowly with peaks corresponding to the spiral structures. The distribution of $A_V$ values in the HST extinction map peaks near $A_V = 0.3-0.4$. Beyond this point, the distribution of $A_V$ values drops like an exponential: $N(A_V) = N_0 \times e^{(-A_V/0.5)}$. The 0.5 value is higher than typical for a spiral galaxy. The outer arms may be tidally distended; the extinction in the corresponding interarm regions is small to an unusually small radius.

To analyse the PPAK IFU data, we take the ratio of a fibre spectrum in the overlap region and the corresponding background fiber spectrum to construct an extinction curve. 
We fit the Cardelli, Clayton and Mathis (CCM) curve to the extinction curve of each fiber element in the overlap region. 
A map of the extinction constructed from PPEX IFU data-cubes shows the same spiral structure of the HST extinction map but the some differences in the distribution of the normalization of the CCM fits ($A_V$). The inferred extinction slopes ($R_V$) maps do not display any structure and a range of values partly due to the sampling effects of the disc by fibers, sometimes due to bad fits, and possibly partly due to some reprocessing of dust grains in the interacting disc.

We compare these findings to our other analysis of an occulting pair with HST and IFU data. In both cases the canonical $R_V=3.1$ is not recovered even though there is enough signal in the extinction curve. We attribute this to mixing opaque and more transparent sections of the disc in each resolution element ($\sim$3" or 0.9 kpc). To illustrate the difficulty of imposing a $R_V=3.1$ law over a section of a spiral disc, we average all spectra and show how a fully grey extinction curve is recovered.
}


\maketitle

\section{\label{s:intro}Introduction}

The distribution of interstellar dust in spiral galaxies is still a dominant unknown in many areas of extra-galactic astrophysics, e.g., models of the spectral energy distribution (SED) of spiral galaxies \citep[e.g.,][Verstappen et al. {\em accepted}, Holwerda et al. {\em in preparation}]{Popescu00, Popescu05rev, Pierini04, Baes03, Baes10a, Bianchi00c, Bianchi07, Bianchi08, Driver08, Jonsson10, Wild11, Holwerda12a} or distance measurements \citep[e.g., SNIa,][]{Jha07, Holwerda08a}. 

Dust is typically studied by either its characteristic far-infrared and sub-millimeter emission or its absorption of starlight in the ultraviolet and optical bands. 
The former approach needs sufficient data to solve for both dust surface density and temperature, requiring a solid detection in multiple bands to break this degeneracy. The latter, in contrast, works even for arbitrarily cold grains, but requires a background light source of known brightness. 

Our knowledge of the geometry of dust in external galaxies has seen great improvements with the {\em Herschel Space Observatory}. These have mapped the temperature gradient in the dusty ISM \citep{Bendo10b,Smith10b, Engelbracht10,Pohlen10, Foyle12}, notably in the spiral arms \citep{Bendo10b} and bulge \citep{Engelbracht10} and the power spectrum of the ISM, gas and dust \citep{Combes12}.  In the case of dwarf galaxies, {\em Spitzer} and {\em 
} results are restricted to a few individual galaxies \citep[e.g.,][]{Hinz06, Hinz12, Hermelo13}, the Herschel Virgo Cluster Survey \citep{Grossi10}, or results of stacked observations \citep{Popescu02b,De-Looze10,Bourne12a}. 
However, the smallest scales can still only be probed in some of the most nearby galaxies, e.g., Andromeda, the Magellanic Clouds or the Milky Way itself. Thus, there is a niche for observations in the optical or ultra-violet to estimate dust geometry in spiral discs independently from sub-mm observations.

To map dust extinction on the smallest scales, one needs a smooth and known background source of light. In the case of a spiral galaxy overlapping a more distant galaxy, the more distant galaxy provides the known background (Figure \ref{f:hst:fov}). Estimating dust extinction and mass from differential photometry in occulting pairs of galaxies was first proposed by \cite{kw92}. Their technique was then applied to all known pairs using ground-based optical images \citep{Andredakis92, Berlind97, kw99a, kw00a}, spectroscopy \citep{kw00b}, and later space-based {\em HST} images \citep[][H09 hereafter]{kw01a, kw01b, Elmegreen01, Holwerda09}. The general results of this technique have been confirmed with the number of distant galaxies seen through spiral discs with HST \citep{Gonzalez98, Gonzalez03, Holwerda05, Holwerda05a,Holwerda05b, Holwerda05c, Holwerda05d,Holwerda05e,Holwerda05f, Holwerda07a,Holwerda07b, Holwerda12c}. More recently, more pairs were found in the SDSS spectroscopic catalog \citep[86 pairs in][]{Holwerda07c}, and in the Galaxy Zoo project \citep[1993 pairs in][]{Keel13}. 

One of these, UGC 3995, an already known interacting pair \citep{Keel86,Keel85,Chatzichristou97,Chatzichristou98,Marziani99,Dultzin-Hacyan99}, was observed with the PPAK IFU as part of the first data-release of the Calar Alto Legacy Integral Field Area Survey \citep[CALIFA,][]{califa} as well as with Hubble Space Telescope as early as 1994.
The wealth of data on this particular pair offers the opportunity to study both small-scales dust extinction features from the Hubble data and the mean and variance of the extinction curve in the foreground spiral from the CALIFA data-cube. The fact that this is a known interacting pair -- similar to the pair studied in \cite{Elmegreen01} -- offers the opportunity to compare the characteristics of the dust disc of a late-type foreground spiral to those of the other superb examples of an overlapping pair presented in \cite{Holwerda09} and \cite{Holwerda13a}, which we suspect is not interacting.

In this paper we aim to map the dust in UGC 3995B\footnote{The galaxies were already identified as UGC 3995 A and B for the background and foreground galaxy. Obviously, UGC 3995B for the foreground galaxy is not optimal terminology in our paper here but we retain it throughout for consistency.} through the HST and CALIFA data and the extinction curves inferred from the CALIFA cubes.
The paper is organized as follows: 
section \ref{s:data} briefly describes the data-products used, 
section \ref{s:obsfx} describes the observational effects in the occulting galaxy technique,  
section \ref{s:analysis} details the analysis and results on both data-sets, 
section \ref{s:disc} is the discussion of the results and 
section \ref{s:concl} lists our conclusions and outlines for future work. We adopt a distance 67 Mpc, from
the mean redshift and ${\rm H}_0 = 71$ km s$^{-1}$ Mpc$^{-1}$.




\section{Data}
\label{s:data}

We use two public data sets for this paper: archival Hubble images, and the two final data-cubes from the first CALIFA data release \citep[DR1, November 2012][]{Husemann13}.

\subsection{HST/WFPC2}
\label{s:hst}

A single WFPC2 {\em F606W} snapshot image, exposed for 500 seconds, of this pair was obtained by the {\em Hubble Space Telescope} in cycle 4 (GO-5479, P.I. M. 
Malkan, as described by \cite{Malkan98} to image the substructure in and around the active center of the background galaxy with the Planetary Camera. Therefore, the PC chip is centered on the nucleus of the background galaxy, UGC 3995A (Figure \ref{f:hst:fov}). Data were obtained from the Hubble Legacy Archive (\url{www.hla.stsci.edu}) and cosmic rays cleaned interactively before analysis.

\subsection{The CALIFA Survey}
\label{s:califa}

The Calar Alto Legacy Integral Field Area Survey \citep[CALIFA,][]{Sanchez12a} is observing a sample of some 600 galaxies in the local Universe using 250 observing nights with the PMAS/PPAK integral field spectrophotometer, mounted on the Calar Alto 3.5 m telescope.
The PPAK instrument offers a combination of a wide field-of-view (2.25 arcmin$^2$) with a high filling factor in one single pointing (65\%), good spatial sampling (1") and wavelength coverage spanning the optical spectrum. The spectra cover the range 3700-7000 \AA\ in two overlapping setups, one in the red (3745-7300 \AA) at a spectral resolution of R$\simeq$850 (the V500 cubes) and one in the blue (3400-4750 \AA) at $R\simeq1650$ (V1200 cubes), where the spectral resolutions quoted are those in the overlapping wavelength range ($\lambda \sim4500$ \AA). 
The targets were selected from the photometric catalog of the Sloan Digital Sky Survey limited in apparent isophotal diameter to fit the PPAK field-of-view. 
A second selection criterion is a  redshift range within $0.005 < z < 0.03$, which ensures that all galaxies can be observed with the same grating settings.
This approach suits our purpose well as it samples the overlap region to  physical scales which could be small enough to resolve the dust structures needed to measure the extinction law \citep[][]{kw01a,Holwerda09}. 
However, the small velocity difference ($\Delta v \sim50$ km/s) between UGC 3995 A and B almost certainly means that the spectral cross-correlation technique pioneered in \cite{kw00b} would require sub-\AA ngstrom spectral sampling. 
CALIFA observations started in summer 2010 and the first data release was in November 2012. Here we use the data cubes publicly available for UGC 3995 from DR1 \citep{Husemann13}.  More occulting galaxy pairs are slated to be observed by CALIFA (\url{http://www.caha.es/CALIFA/public_html/}).

\subsubsection{Fibre-transmission and Sky-subtraction}

The CALIFA data-reduction is described in detail by \cite{Sanchez12a}. We focus on the relative fibre spectral transmission and the sky-subtraction because these are the most relevant for our fibre-to-fibre comparison to obtain extinction curves. 

The relative transmission of the fibres is calibrated with a lamp spectrum. The accuracy of the relative transmission is better than a few percent in most of the covered fibres.
The initial {\em absolute} spectrophotometric calibration was relatively poor (24\% on average nights, and poorer in non-photometric condition) but the final product is accurate to 
$\sim$8\% after a re-calibration using the SDSS imaging of these galaxies \citep[see \S 5.7 in][]{Sanchez12a}. 
The relative transmission affects the measured {\em slope} of the extinction curve ($R_V$) and the absolute calibration the normalization of this curve ($A_V$).

The subtraction of the night-sky spectrum from the data is an issue in many IFU observations of nearby galaxies, mostly due to the relatively small
field-of-view and the resulting lack of a pure night-sky spectrum, uncontaminated by source flux. We note the VIMOS/IFU observations in 
\cite{Holwerda13} were an exception, because that pair was small enough on the sky. By design, the apparent size of the galaxies in CALIFA fit within the FoV of the central bundle of the PPAK IFU.
The PPAK IFU has 36 dedicated sky fibers, positioned such that they are free from any source contamination. With these clean sky-fibres, subtraction is straightforward and not a main source of uncertainty: the sky-subtraction has an accuracy of 1-5\%. 

\subsubsection{Spatial Resolution}

The PPAK has a relatively coarse sampling of the field-of-view. To mitigate this, a well-understood three-pointing dithering strategy was adopted for the CALIFA survey \citep{Sanchez12a}. Night-specific measurements of the seeing conditions are presented for DR1 observations in \cite{Husemann13}: typical seeing was 0\farcs9 and thus is not the limiting factor in the spatial resolution of the CALIFA data cube. The image reconstruction algorithm of the dithered observations resulted in a typical spatial resolution of 3" FWHM (similar to the fibre diameter of 2\farcs7), as verified with bright stars in the FOV. We adopt a fiber size of 1", and resolution 3", verified by an examination of the star in Figure \ref{f:cal:fov} in both cubes.
Improvements in the spatial sampling are expected in future data-releases of CALIFA.

\section{Observational Effects}
\label{s:obsfx}

There are three observational effects to consider in extinction measurement using an occulting pair of galaxies: intrinsic asymmetry of either galaxy, light scattered into the line-of-sight and the effects of physical sampling of the foreground ISM. We want to point them out to the reader and briefly review them.

\subsection{Symmetry of both Galaxies}
\label{s:asymmetry}

The central assumption in our analysis of the HST and IFU data is that both galaxies are rotationally symmetric. This assumption of symmetry must especially hold for the background galaxy. For this reason, we would prefer an elliptical galaxy as a background. However, in this case it is a grand-design spiral. For this reason, observations in the red are preferred as spirals appear more regular without much structure from star-formation which often breaks symmetry on small scales. 

In principle, the dark dust structure visible in the HST image (Figures \ref{f:hst:fov}, and \ref{f:hst:Avmap}) and the attenuation evident in the IFU data 
could be part of the background galaxy. That would uncharacteristically break symmetry for the background galaxy, and secondly, the general morphology of the dusty structures strongly suggests that they belong to the foreground disc (orientation, implied winding angle) and not the background galaxy. The dust lies on the extrapolated spiral arm pattern of the foreground galaxy, not the background galaxy's.
However, the deviations from symmetry due to either local star-formation or extinction in the background galaxy are likely the dominant uncertainty in the extinction measurement.

\subsection{Scattering}
\label{s:scattering}

Because the galaxies are not point sources but extended objects, the relation between attenuation and wavelength may include effects of background galaxy light scattered {\em into} the line-of-sight by dust clouds in the foreground disc. This effect is treated in detail in the appendix of \cite{kw00a}. There are two extended sources of scattered light, the foreground and background galaxy. 

The foreground galaxy contribution is removed in the HST analysis through subtraction. 
The contribution of light from the foreground galaxy scattered within the foreground itself is removed by the symmetry in our analysis: the subtraction of the foreground flux in equation \ref{eq:T} leaves no foreground scattered light.
This leaves the background galaxy as the main source of scattered light in our measurements. 
The effect of scattering is greatest in regions of modest optical depth ($\tau\sim1$) and brightest background source (close to the background galaxy's nucleus). Any opaque regions are self-absorbing and this effect does not matter. 

\cite{kw00a} examine the contribution of scattered light into a line-of-sight as a function of angular and radial separation of the two galaxies. If the galaxies are sufficiently far apart, the scattering angle becomes negligible. They express the scattering percentage as a function of the ratio between radial extent and separation of both discs. Given the small velocity and distance separation between UGC 3995A and UGC3995B ($\Delta \sim 50$ km/s velocity separation\footnote{An interacting system could have a $\Delta v = 300$ km/s and yet be spatially coincident. Unfortunately, more accurate distances for both galaxies are not available.}), one would expect an effect of several percent of observed flux in the overlap. The main effects would be to lower the contrast between opaque and transparent regions in the HST extinction map and, because scattering affects blue light more than red, a greyer observed extinction curve.

\subsection{Sampling}
\label{s:sampling}

In H09, we already noted the effect of spatial sampling on the observed extinction law. By mixing opaque and transparent regions in the same resolution element, the effectively observed extinction curve becomes much less wavelength dependent, i.e., a ``grey" extinction law \citep[$R_V << 3.1$, see][]{Calzetti94}. This has little effect on the HST extinction map as it possesses exquisite spatial resolution and no colour information. However, this can be an important observational effect in the CALIFA results. We return to this in the discussion of the results below.

\section{Analysis}
\label{s:analysis}

Analysis of the HST and CALIFA data is discussed separately with a third section comparing the results.

\begin{figure}
\begin{center}
\includegraphics[width=0.5\textwidth]{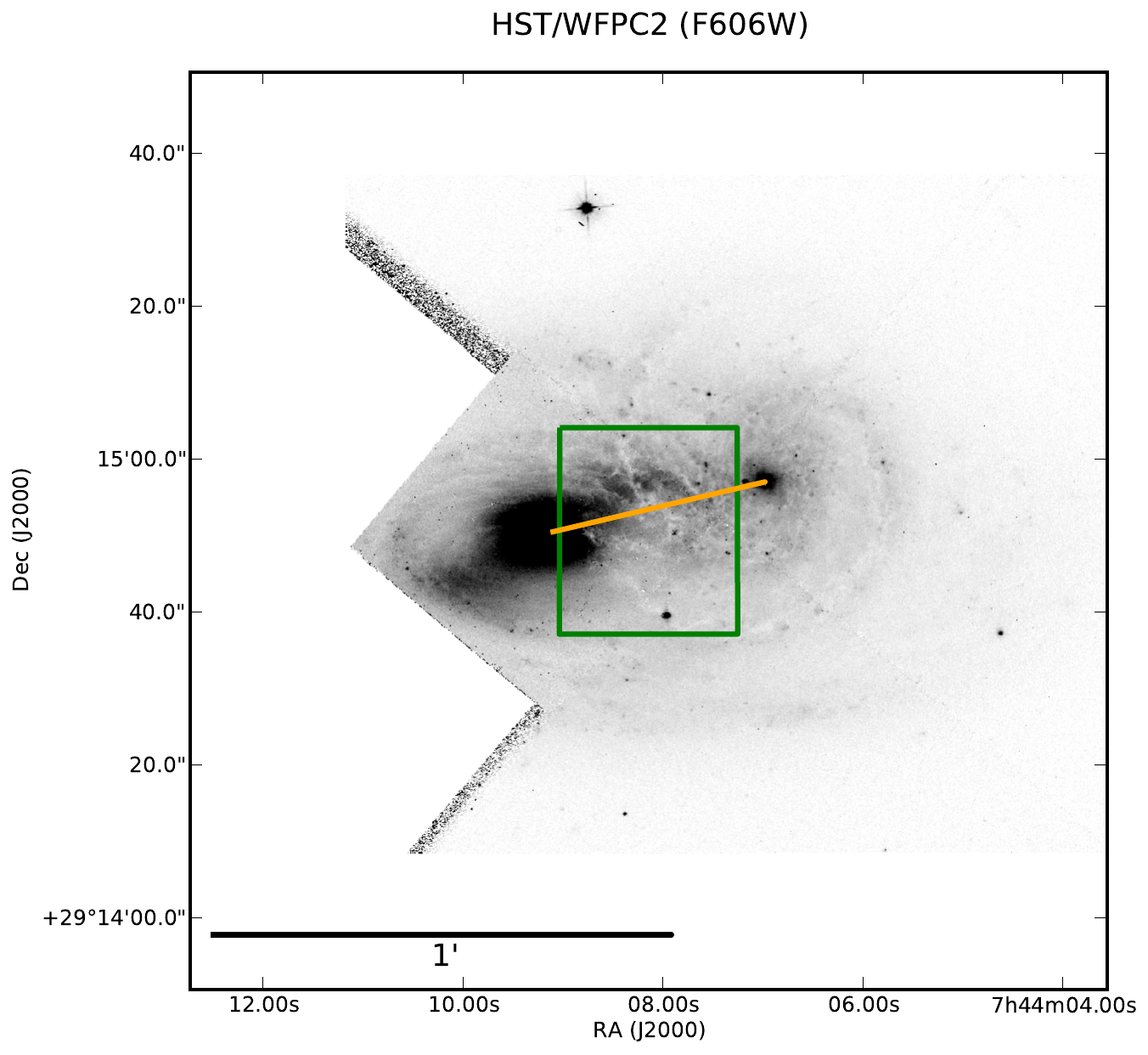}
\caption{An inverse greyscale plot of the HST/WFPC2 $F606W$ image with the overlap aperture (green box). UGC 3995B is the foreground spiral on the right, UGC 3995A is the bright background spiral on the left. The nucleus of UGC 3995A was the target of GO-5479 with the Planetary Camera part of the WFPC2. The orange line shows over which the radial profile in Figure \ref{f:hst:T} was taken.} 
\label{f:hst:fov}
\end{center}
\end{figure}

\begin{figure}
\begin{center}
\includegraphics[width=0.5\textwidth]{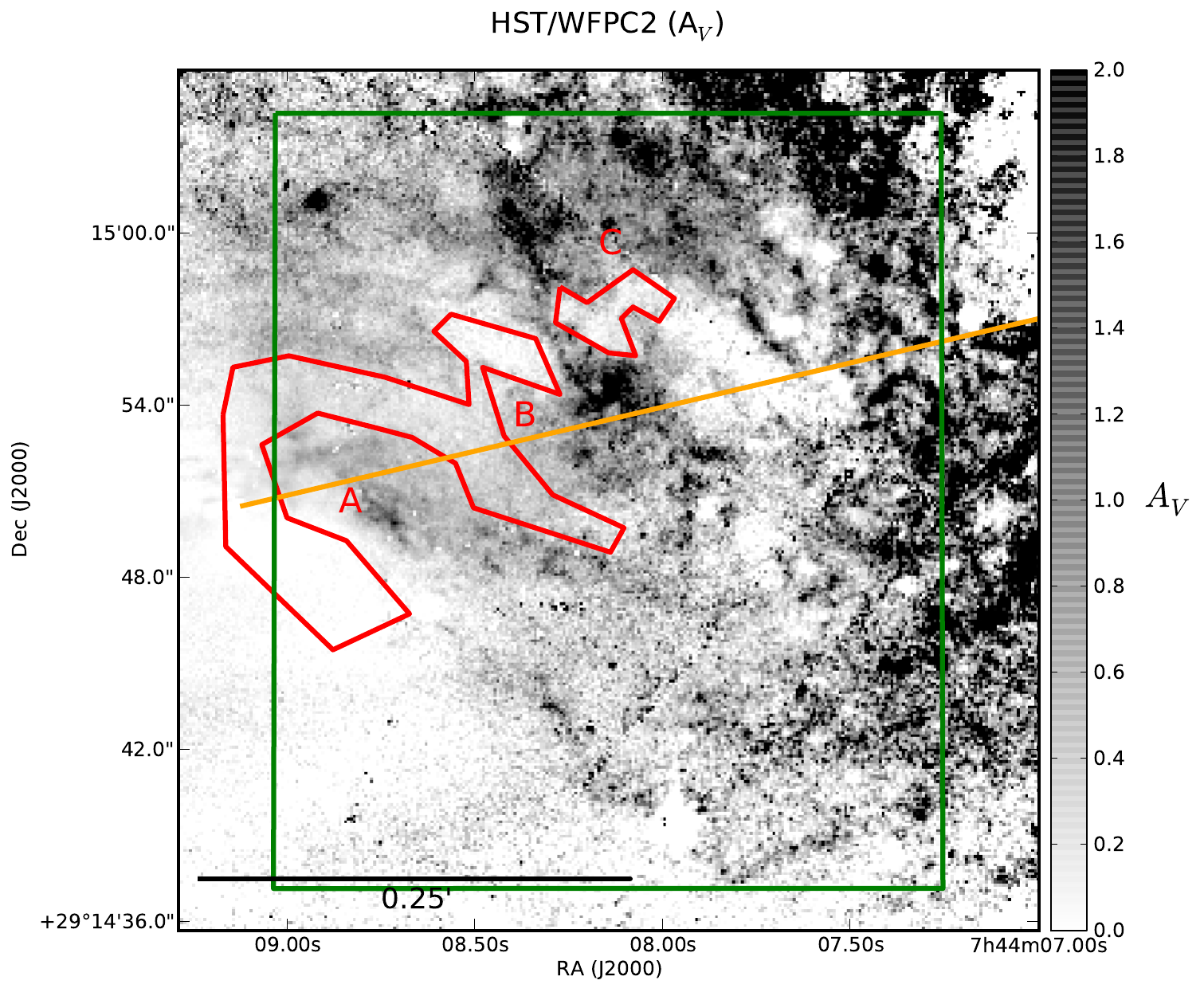}
\caption{The extinction map, corresponding approximately to the overlap region marked in Figure \ref{f:hst:fov}, in which we identified the inter-arm regions of UGC 3995B by eye (red polygon), surrounding complex A and B in Figure \ref{f:hst:Avmap}. The orange line shows where the transmission curve from the center of UGC 3995A to the center of UGC 3995B, shown in Figure \ref{f:hst:T}, was taken. }
\label{f:hst:Avmap}
\end{center}
\end{figure}

\begin{figure}
\begin{center}
\includegraphics[width=0.5\textwidth]{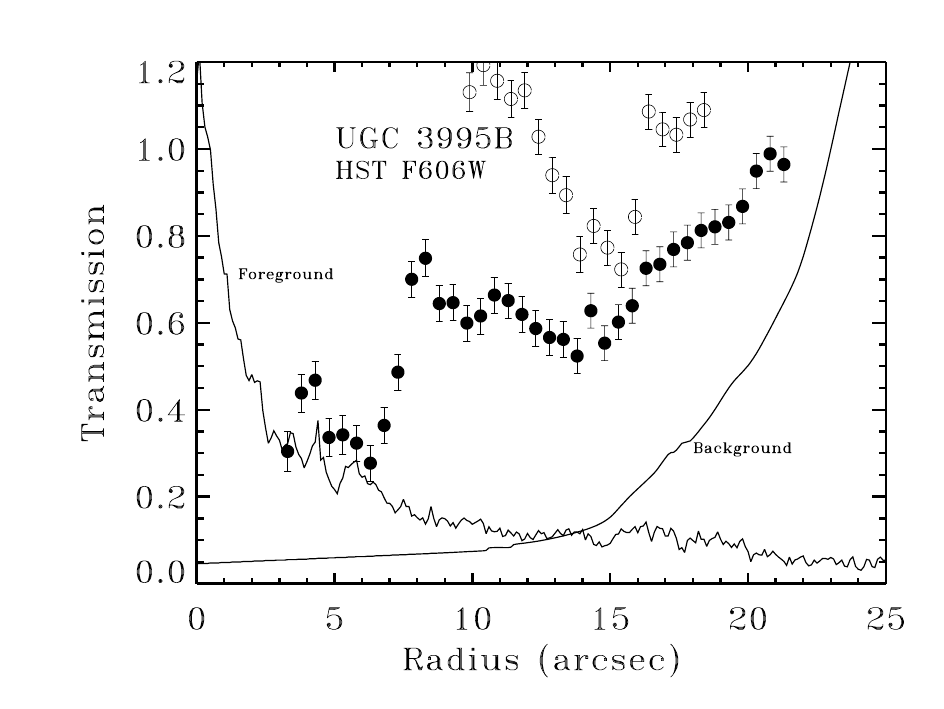}
\caption{A radial plot of the transmission $T$, which is the fraction of incident light reaching us through the foreground galaxy. 
Solid points are the averaged radial profile of transmission, with error bars at each radius from the scatter in measurements (error of the
mean in each case). Open symbols show inter-arm regions as identified in the transmission map (Figure \ref{f:hst:Avmap}), in regions of high expected S/N. Solid lines are the intensity of light from foreground (UGC 3995B) and background (UGC 3995A) galaxies along the line between both nuclei (the orange line in Figure \ref{f:hst:fov}), to the same but arbitrary scale.}
\label{f:hst:T}
\end{center}
\end{figure}


\begin{figure}
\begin{center}
\includegraphics[width=0.5\textwidth]{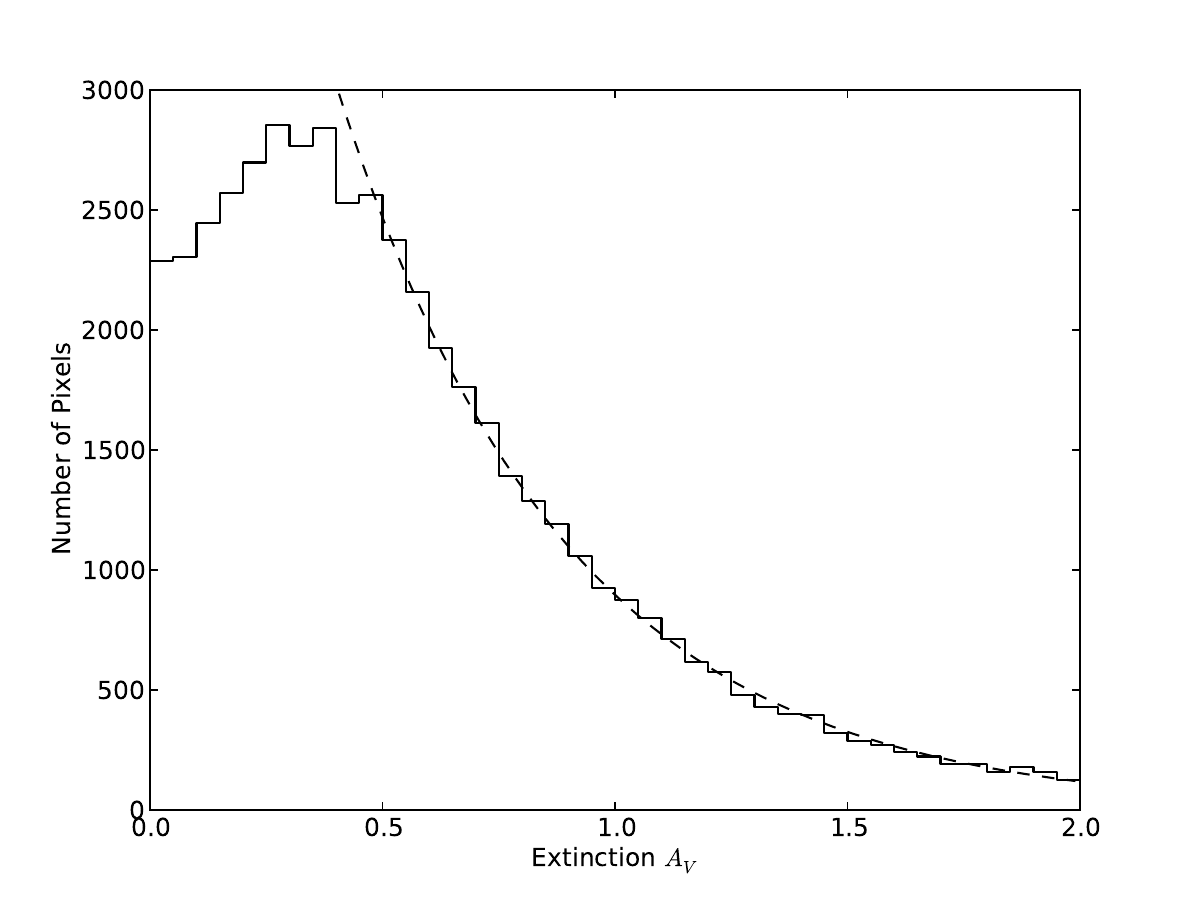}
\caption{The distribution of $A_V$ values in the green aperture in Figure \ref{f:hst:fov} or \ref{f:hst:Avmap}. The dashed line is an exponential drop fit to the values beyond $A_V =0.5$. Above this value, the distribution of pixels with  is well described by $N = N_0 \times e^{(-A_V/0.5)}$ with $N_0=6799$ pixels. }
\label{f:hst:Avhist}
\end{center}
\end{figure}

\begin{figure}
\begin{center}
\includegraphics[width=0.5\textwidth]{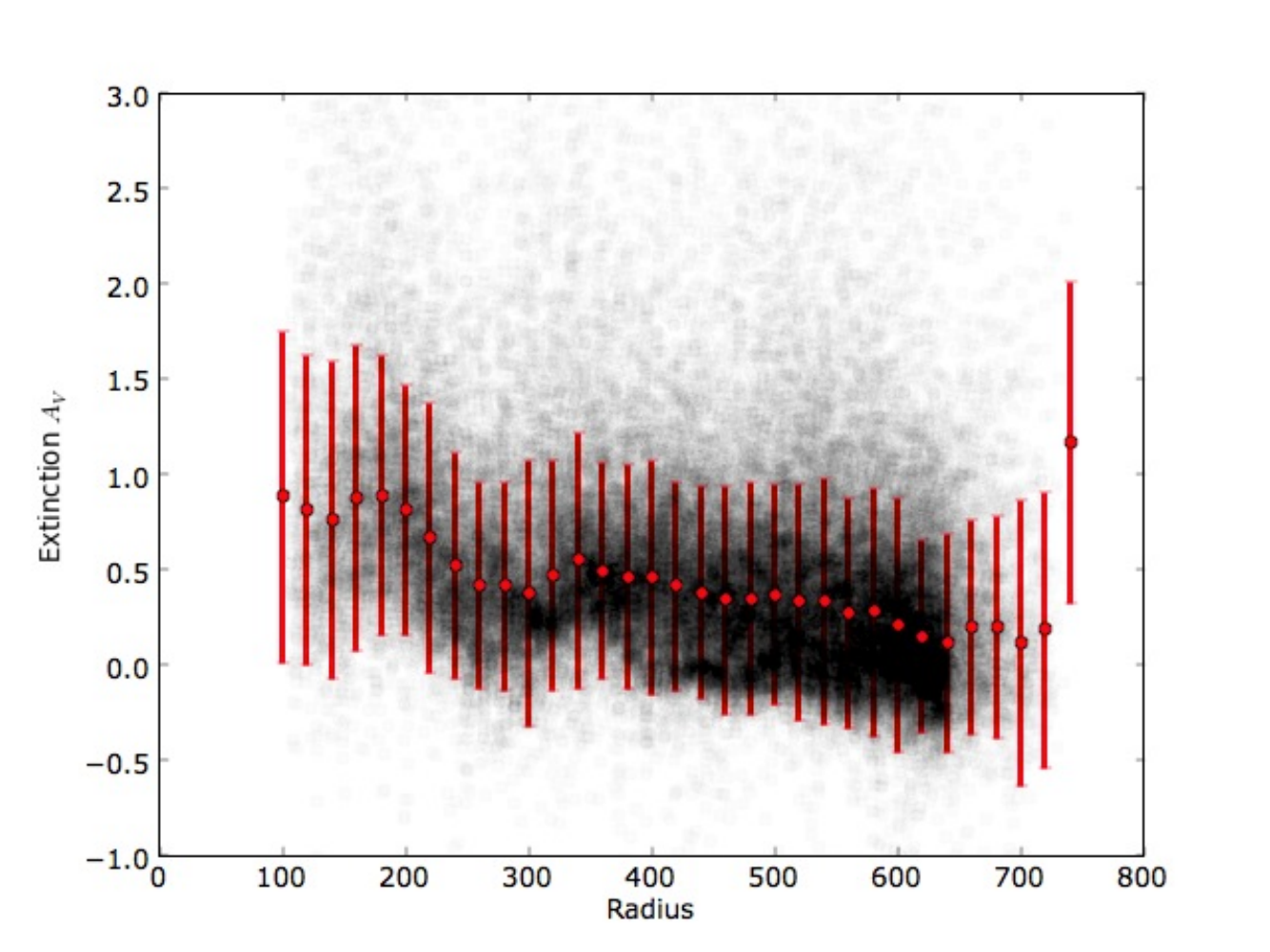}
\caption{The projected distance from the center of UGC 3995B in arcseconds versus the extinction ($A_V$). Gray points are individual pixel values in the aperture in Figure \ref{f:hst:Avmap}, the red points and error bars are the median and rms of these values.}
\label{f:hst:RAv}
\end{center}
\end{figure}

\subsection{HST Data Analysis}
\label{s:hst:analysis}

The analysis of the HST/WFPC2 image (Figure \ref{f:hst:fov}) extends the work by \cite{Marziani99} with a more detailed procedure. 
We use elliptical isophote fits to the unobscured part of UGC 3995A to model its light distribution, supplementing the HST 
image with the SDSS $r$ data in the gaps around the PC CCD, and use symmetry of UGC 3995B to correct for the contribution 
of foreground light. We found this approach the most reliable in H09 for the application of this method to HST data.
The geometry of the UGC 3995 pair is especially favorable, in that the outer arms of UGC 3995B are projected in 
front of the bright, strongly bulge-dominated parts of UGC 3995A. Departures from the symmetry assumptions will be manifested 
differently for the two galaxies, largely following the spiral patterns of the respective galaxies. For the radial extinction profile, averaging 
azimuthally (or along the arm pitch angle) has the effect of marginalizing over the nuisance parameter of location along the background arm. 

The transmission $T$, which is the fraction of incident light reaching us through the foreground galaxy, is well behaved to errors from either galaxy model, i.e., it depends on either linearly, and we use this quantity to map extinction. As in \cite{kw92} and H09, if $B$ and $F$ are the intensities 
of background and foreground galaxies at each point estimated from symmetry or modeling, $T$ may be estimated from the observed intensity $I$ following:

\begin{equation}
\label{eq:T}
T = {{I - F} \over { B}},
\end{equation}
\noindent $T$ is related to the extinction of the dust in the foreground disc as:
\begin{equation}
\label{eq:Av}
A_V = -1.085 \times ln (T)  =  -1.085 \times ln \left({{I - F} \over { B}}\right)
\end{equation}

Figure \ref{f:hst:Avmap} shows the extinction ($A_V$) map, based on the WFPC2 field. There is clear noise at the edges of the FOV as the background galaxy contribution is the lowest. Several foreground stars remain as artifacts. However, in the aperture (green box), there is sufficient signal-to-noise for a meaningful extinction measure ($A_V$). Several spiral arms are evident in extinction (regions A and B in Figure \ref{f:hst:Avmap}, also visible in Figure \ref{f:hst:fov}) with a myriad of substructure in each arm.

Figure \ref{f:hst:T} shows the averaged radial profile of transmission from the HST image, with error bars at each radius from the scatter in measurements (error of the mean in each case); open symbols show inter-arm regions as identified in the transmission map, in regions of high expected S/N. This is an unusually complete profile in radius from the overlap technique, comparable to that of NGC 3314A shown by \cite{kw01b}.
The curves in Figure \ref{f:hst:T} show the intensity of light from foreground and background galaxies along the line between nuclei (orange line in  \ref{f:hst:Avmap}), as evaluated from our modeling of symmetric foreground and ellipse-fit background. This provides a guide to how much scatter is introduced by structure in the foreground galaxy, which is important where it is much brighter than the background. These intensity curves of the foreground and background galaxy are to the same (arbitrary) scale.

The dust lanes in the outer arms are systematically offset to greater radius with respect to the stellar light associated with the spiral arm (Figure \ref{f:hst:T}); dust leads stellar spiral structure.
The reliable identification of individual extinction features becomes progressively less certain toward the center of UGC 3995B, since the background light is weaker and the amplitude of contaminating foreground features becomes larger. Conversely, the dust in several of the outer arms is particularly well and reliably mapped, e.g., complexes A and B in Figure \ref{f:hst:Avmap}. 
In Figure \ref{f:hst:T}, we note a fairly sharp edge to the occurrence of these dust lanes at 21" (6.8 kpc), about 1.6 disc scale lengths based on fitting the radial profile of the non-backlit part of the galaxy (orange line in Figure \ref{f:hst:Avmap}).

In Figure \ref{f:hst:Avmap} the inter-arm regions can be clearly distinguished in the outer disc of UGC 3558B, partly by their lower extinction. %
Their transmission are the open points in  Figure \ref{f:hst:T}. Inter-arm regions are mostly transparent but block up to 20\% of the light of the background galaxy. Some inter-arm measurements show transmission greater than unity, the result of asymmetry in the background galaxy's light distribution.

The arm geometry seen in absorption agrees with the visual impression that UGC 3995B is viewed essentially face-on, so we make no additional corrections for projection effects in estimating optical depths. At least on the eastern (backlit) side of the disc, the dust arms lie systematically outside the ridge line of stellar arms.

Figure \ref{f:hst:Avhist} shows the histogram of the overlap aperture in Figure \ref{f:hst:Avmap}. The distribution peaks at $A_V \sim 0.25$, which is unusual for the outer disc of a spiral galaxy. In the pair we analysed in H09 and \cite{Holwerda13a}, the distribution gradually declines from a maximum at $A_V=0$. We suspect that this other pair is not interacting (the respective redshifts in that case imply a separation of $\sim$4 Mpc, still within the velocity range of bound pairs, so their interaction status remains ambiguous). The redshift separation between UGC 3995A and B is only $\sim 50$ km s$^{-1}$, so they are certainly tidally interacting. A recent or ongoing tidal interaction might be responsible for the redistribution into denser clouds through compression, shocks and triggered collapse.  The offset of the peak from $A_V=0$ and the high value of the exponential decline in Figure \ref{f:hst:Avhist} could be related to the very transparent interarm regions noted above, if the outer arms are at least partly tidally shaped; tidal interaction removed diffuse inter-arm dust into dense concentrations in the arms. 

We fit an exponential to the decline of the distribution in Figure \ref{f:hst:Avhist} beyond $A_V=0.5$. We find an exponential drop of 0.5, i.e, the number of pixels with value $A_V$ can be well described as $N(A_V) = N_0 \times e^{(-A_V/0.5)}$. This value is higher than we found in H09 and it is not typical of the exponential distribution used as a prior, for example, in SN\,Ia lightcurve fits (typically the drop-off is 0.3). 

Figure \ref{f:hst:RAv} shows the radial distribution of all the extinction measurements in the overlap aperture (green rectangle in Figure \ref{f:hst:Avmap}) and their median. The extinction gradually decreases with distance from the center of the foreground galaxy UGC 3995B. Two bumps in the mean value point to the spiral structures visible in Figures  \ref{f:hst:fov}, and \ref{f:hst:Avmap}.


\begin{figure*}
\begin{center}
\includegraphics[width=0.49\textwidth]{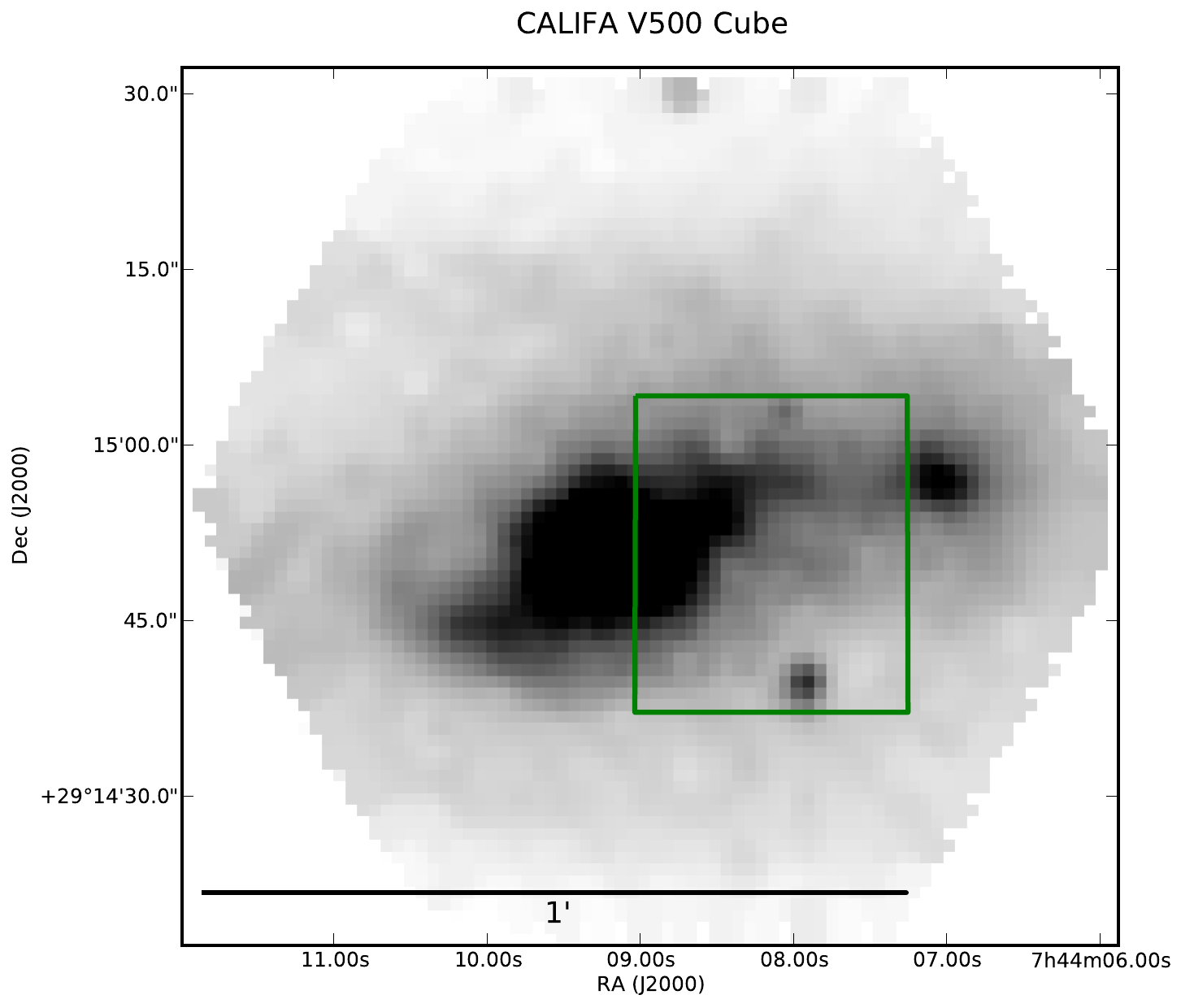}
\includegraphics[width=0.49\textwidth]{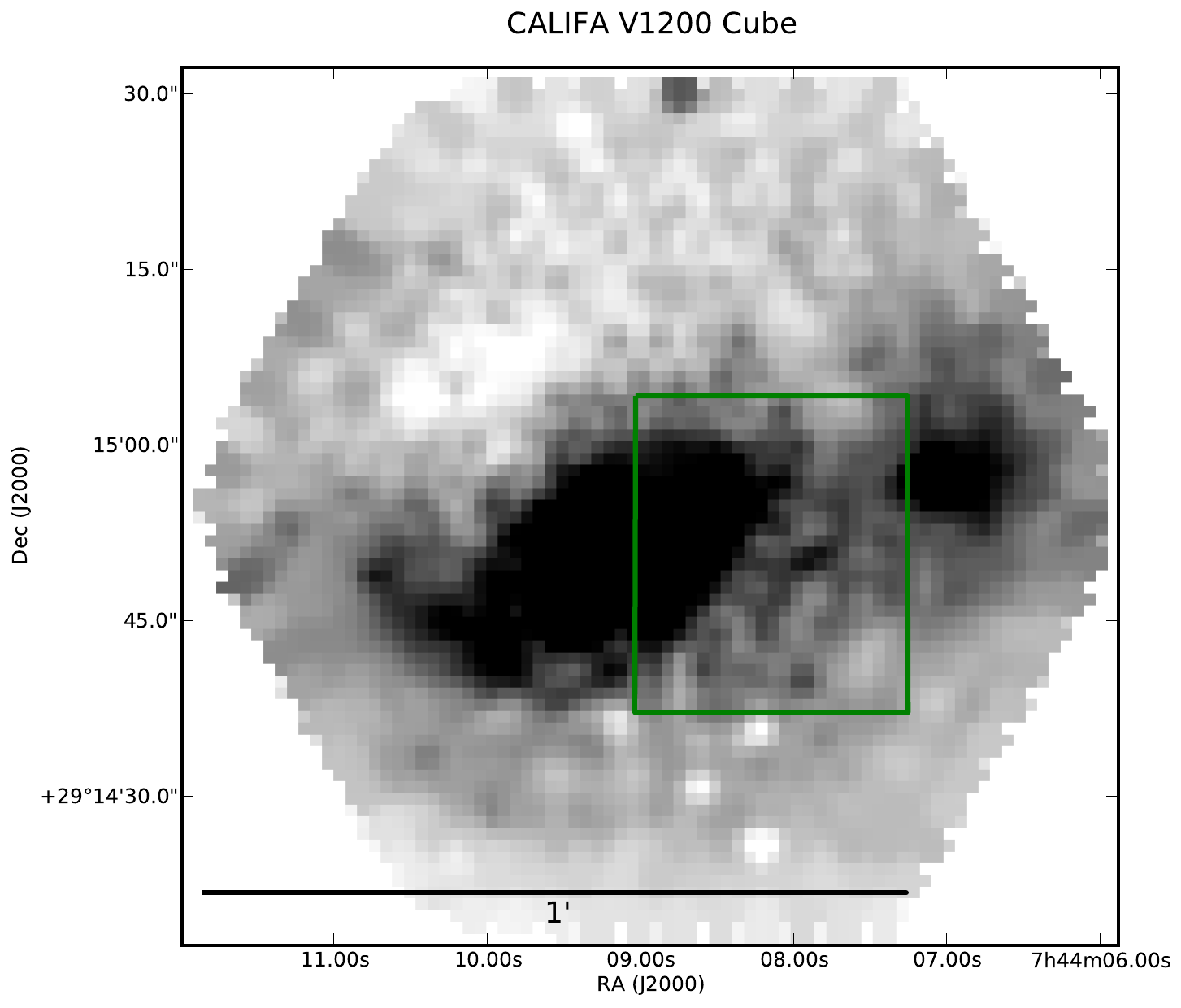}
\caption{The summed PPAK data-cubes from the CALIFA, the V500 (left) and V1200 (right).}
\label{f:cal:fov}
\end{center}
\end{figure*}

\begin{figure*}
\begin{center}
\includegraphics[width=0.49\textwidth]{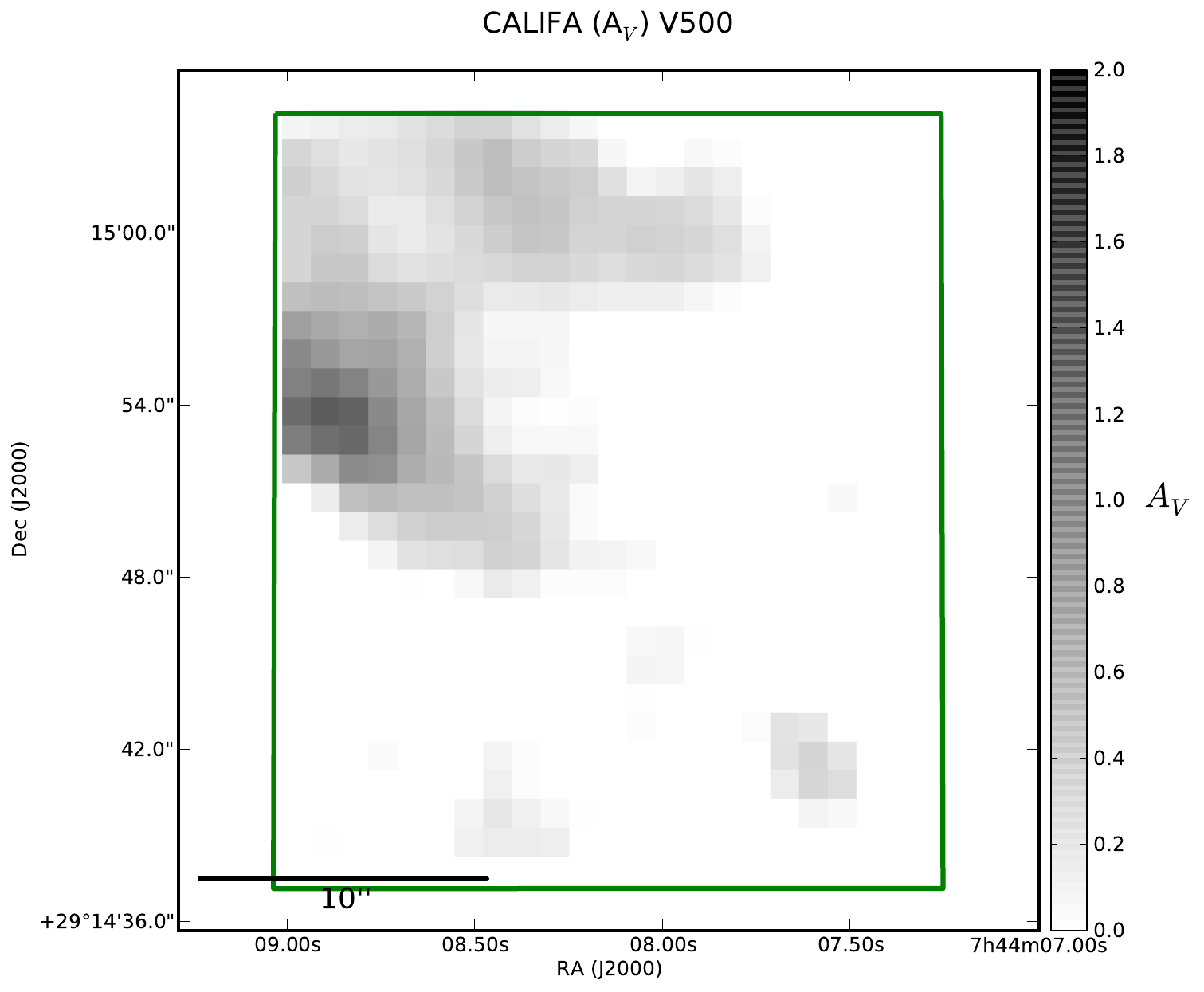}
\includegraphics[width=0.49\textwidth]{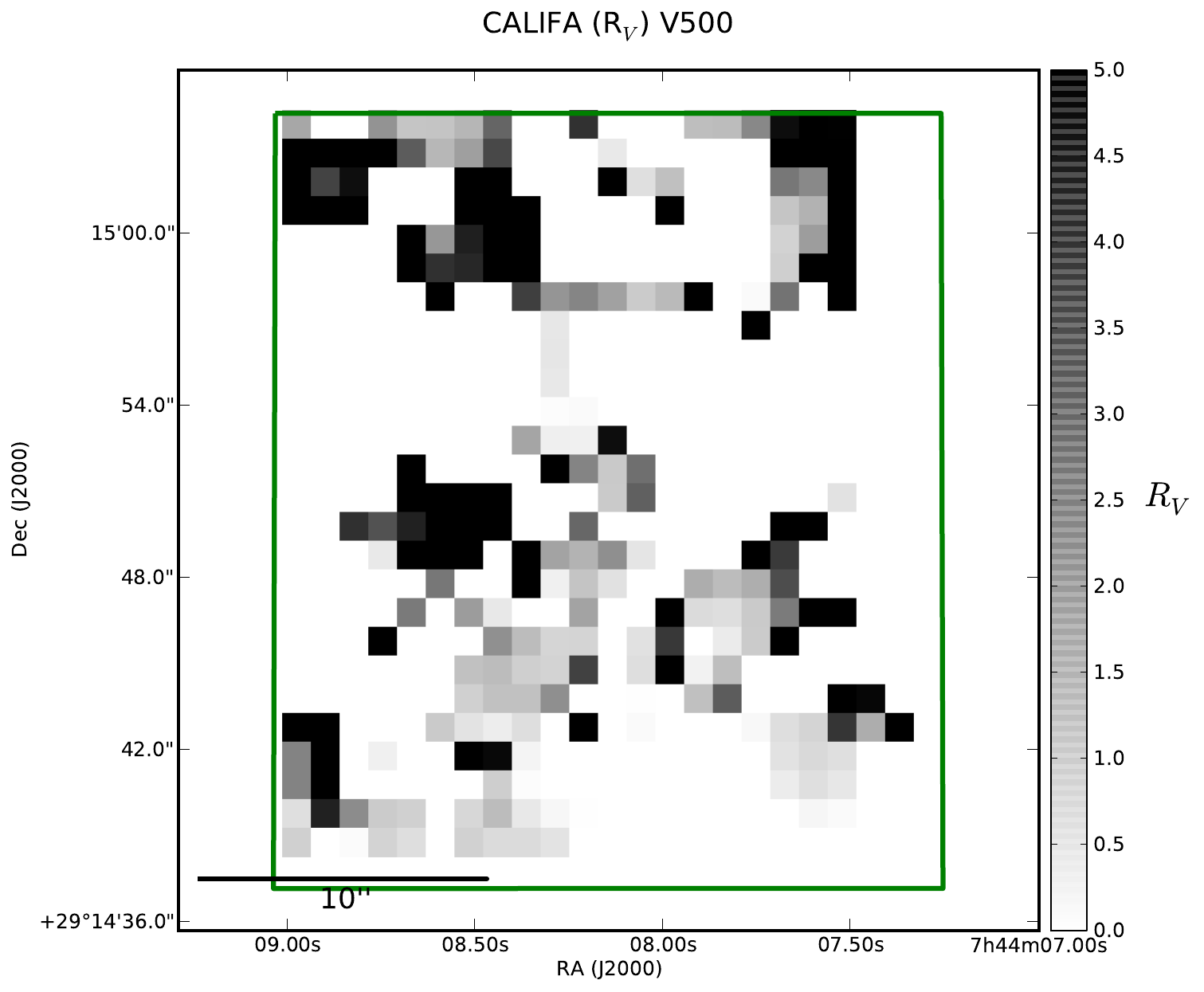}
\includegraphics[width=0.49\textwidth]{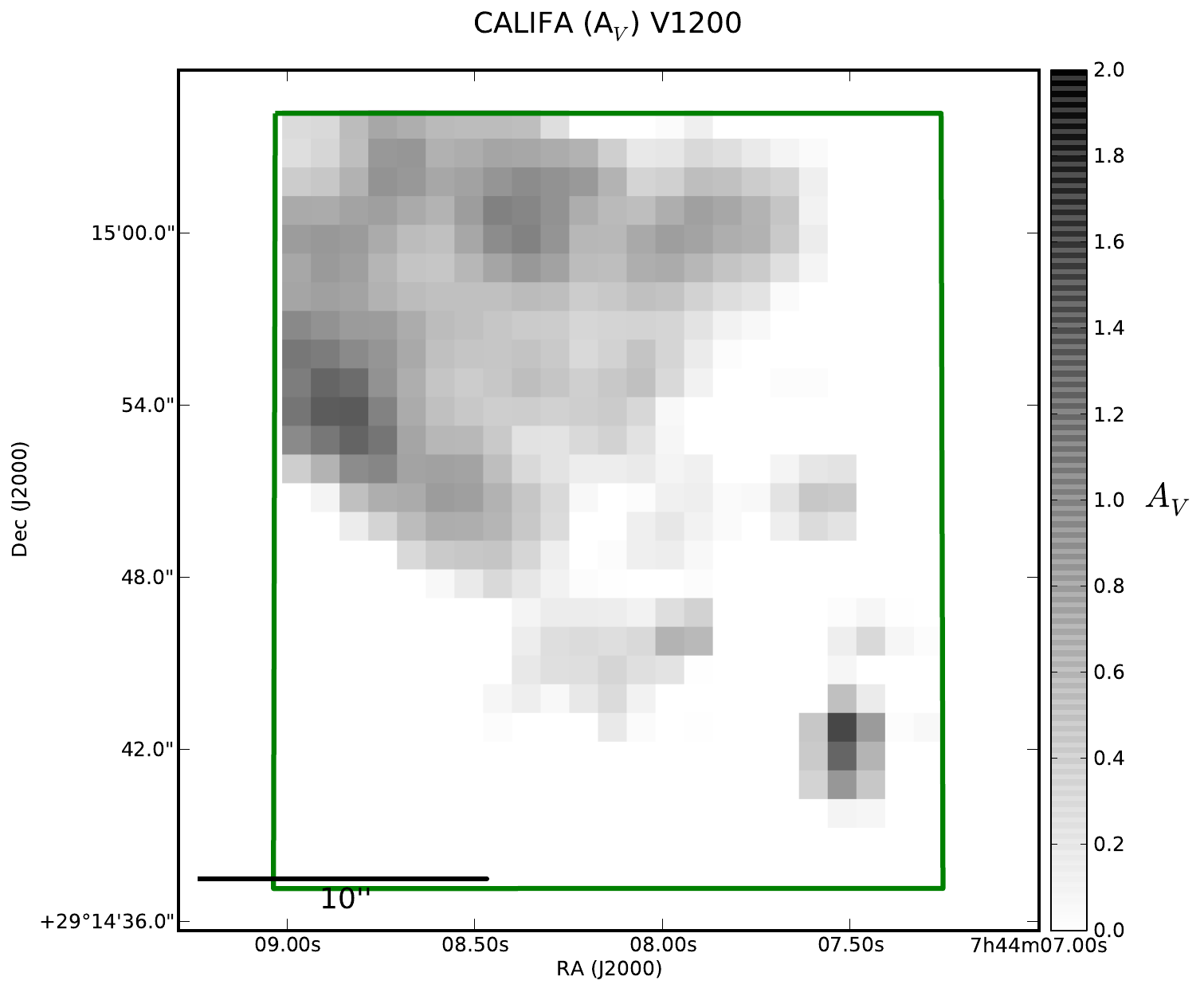}
\includegraphics[width=0.49\textwidth]{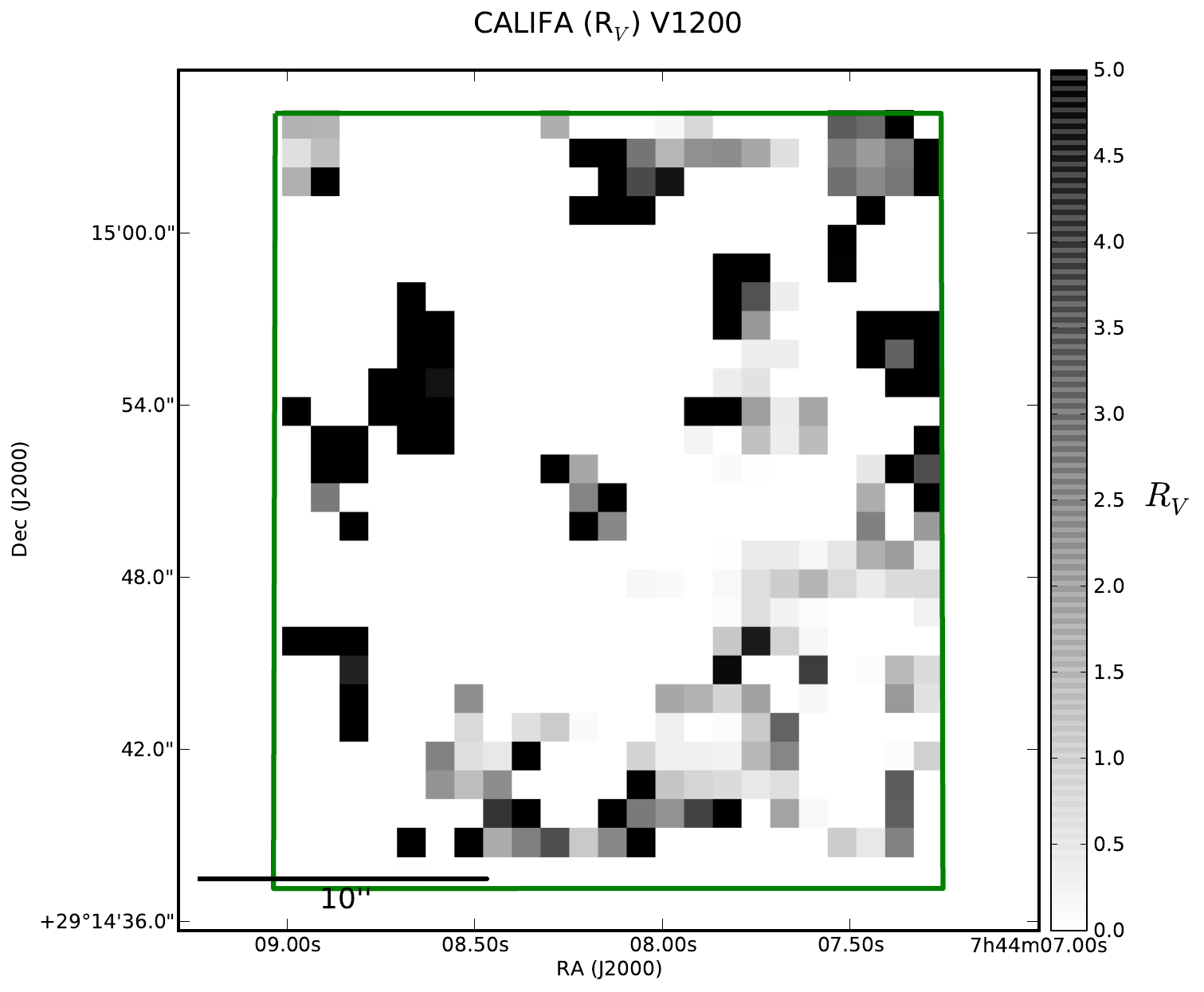}
\caption{Maps of the normalization ($A_V$) and slope ($R_V$) of the CCM fit to the fiber extinction curves ($\rm \tau = -1.086 \times ln(OL/BG)$) from the V500 and the V1200 cubes. The normalization follows the extinction map in Figure \ref{f:hst:Avmap} but the slope, $R_V$ is unrelated to the extinction in the HST image. }
\label{f:cal:ccm}
\end{center}
\end{figure*}

\begin{figure*}
\begin{center}
\includegraphics[width=0.49\textwidth]{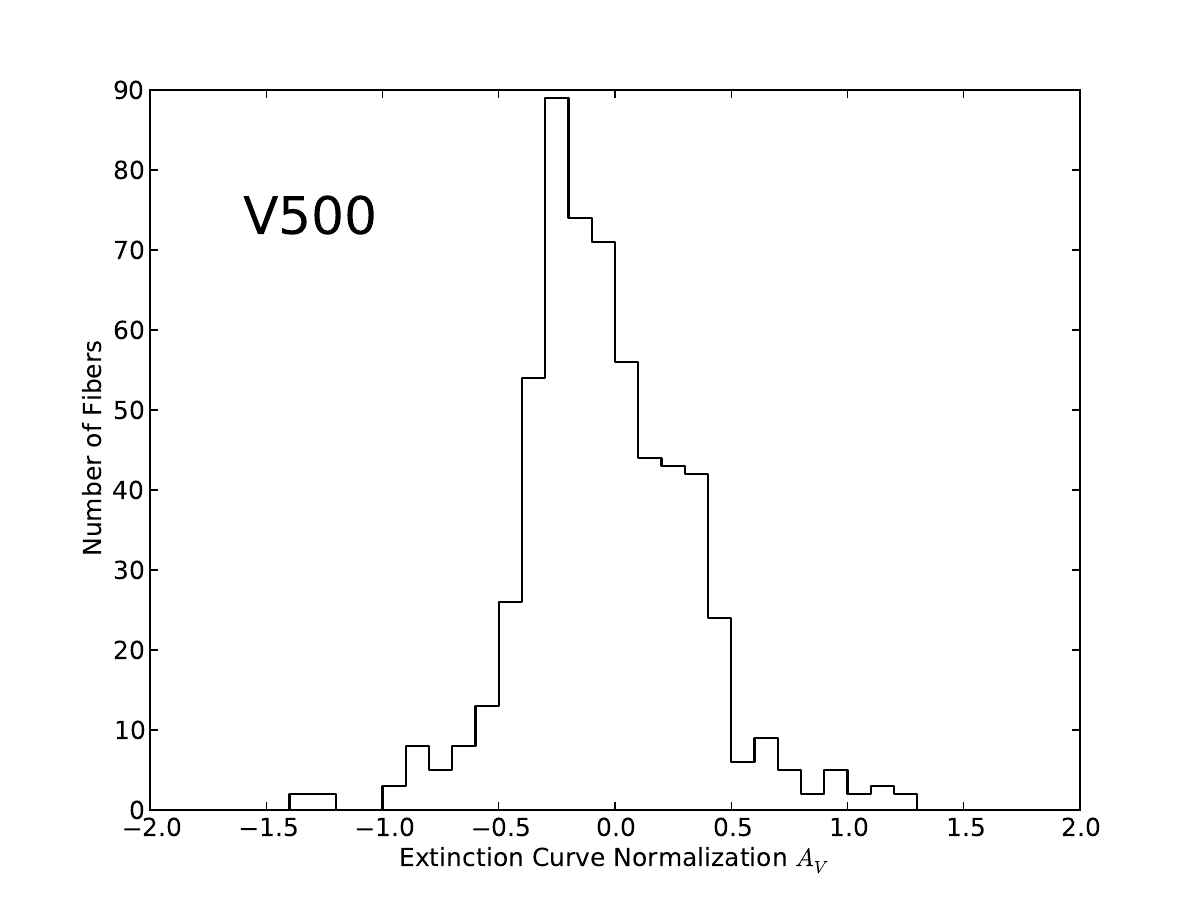}
\includegraphics[width=0.49\textwidth]{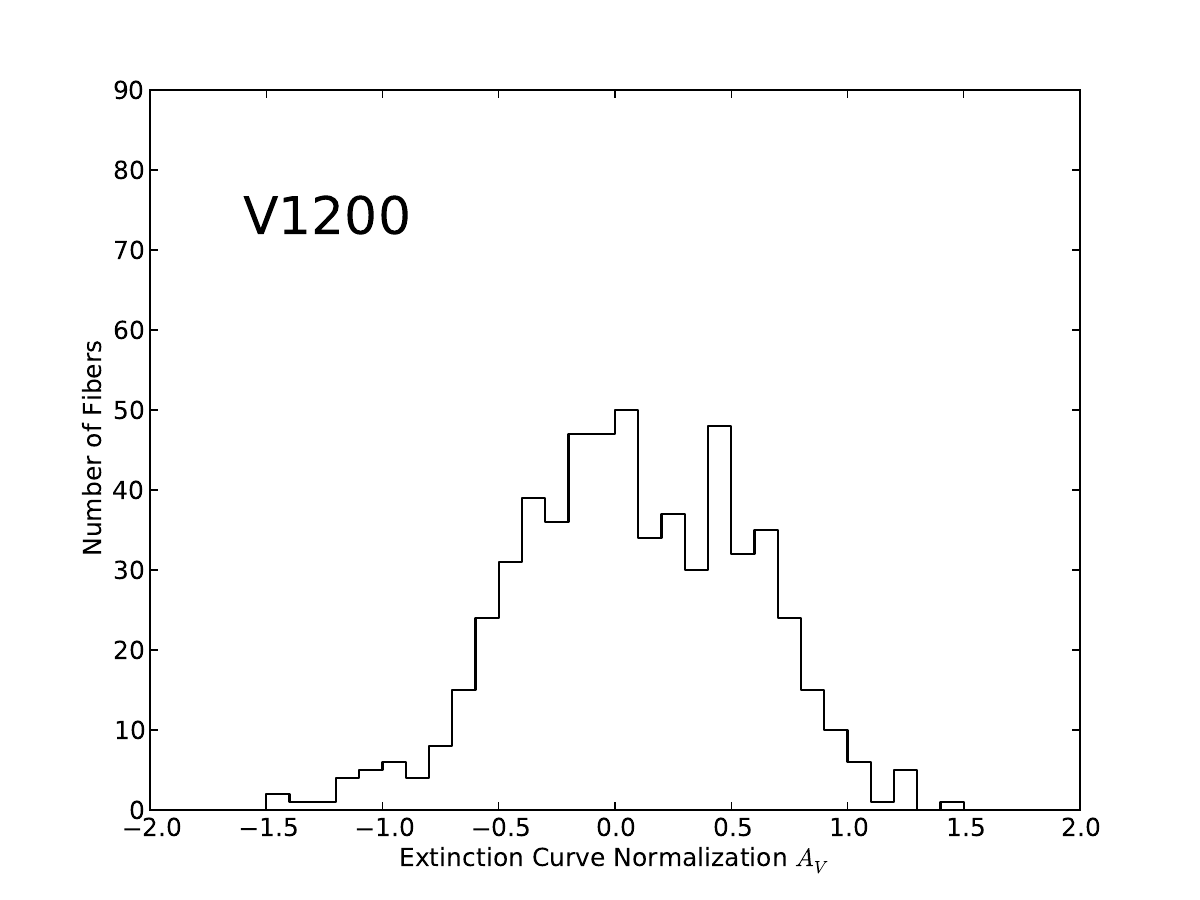}
\caption{The distribution of normalization ($A_V$) values in the aperture region (Figure \ref{f:cal:fov}) for the CCM fits to the extinction curves in the fibers in the V500 (left) and V1200 (right) cube. Because the wavelength range is different for each cube, a different normalization can be expected.
}
\label{f:cal:Avhist}
\end{center}
\end{figure*}

\begin{figure*}
\begin{center}
\includegraphics[width=0.49\textwidth]{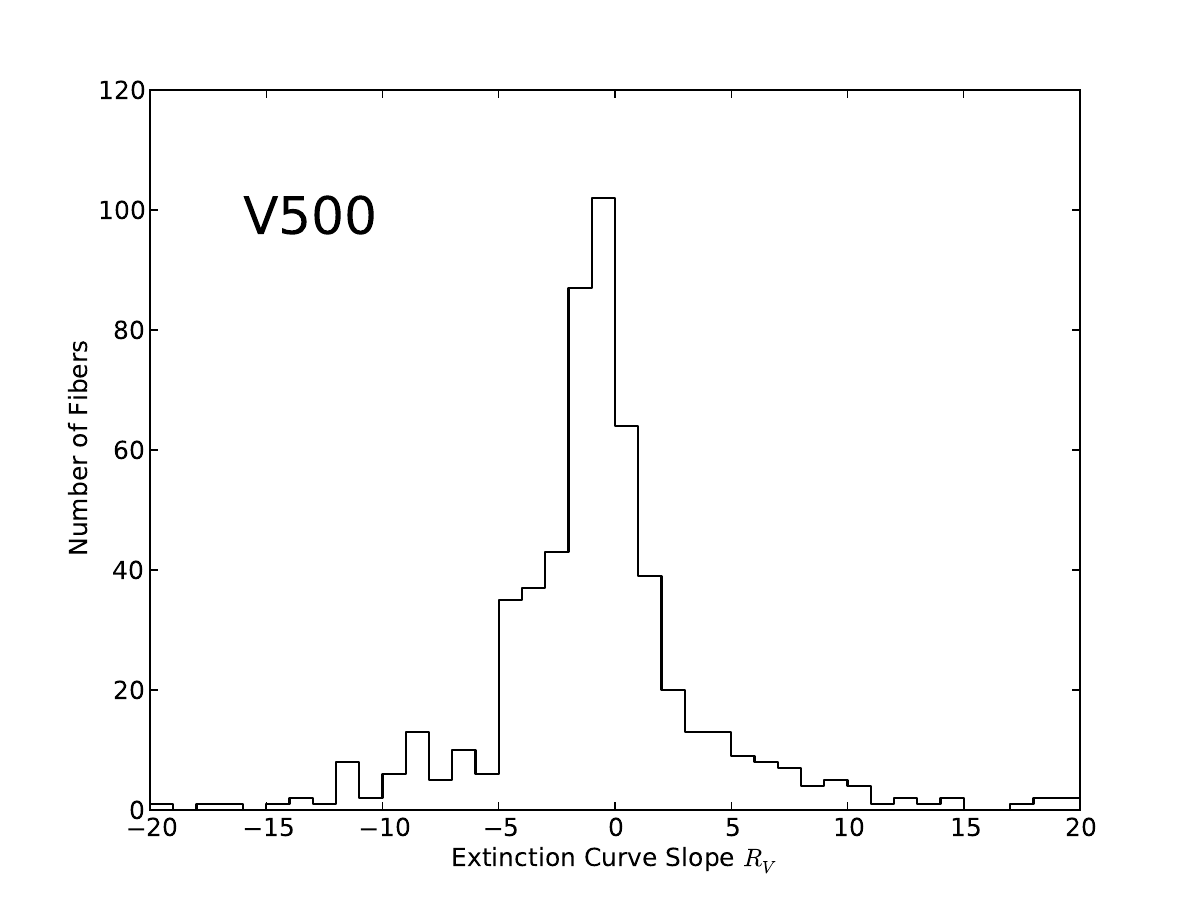}
\includegraphics[width=0.49\textwidth]{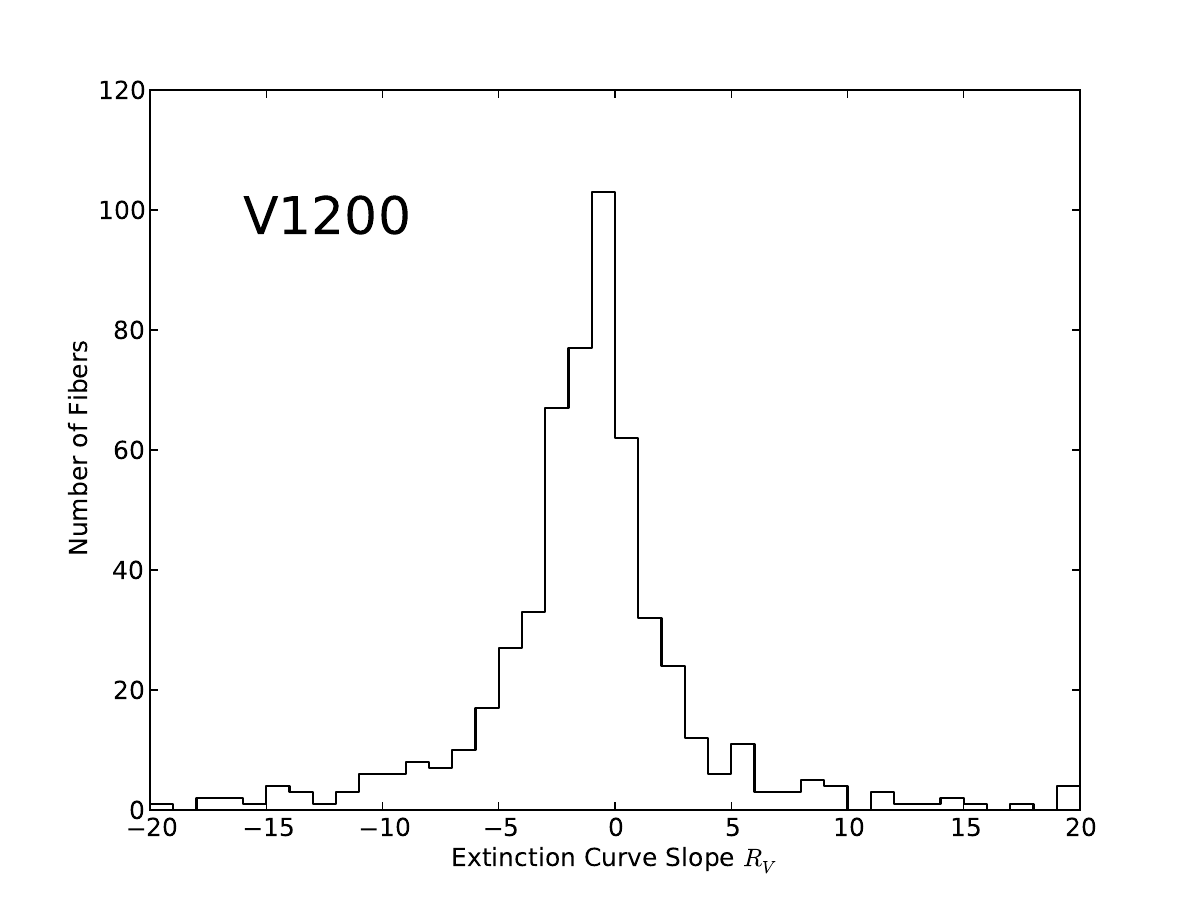}
\caption{The distribution of the slope ($R_V$) values in the aperture region (Figure \ref{f:cal:fov}) for the CCM fits to the extinction curves in the fibers in the V500 (left) and V1200 (right) cube. 
}
\label{f:cal:Rvhist}
\end{center}
\end{figure*}

\begin{figure}
\begin{center}
\includegraphics[width=0.49\textwidth]{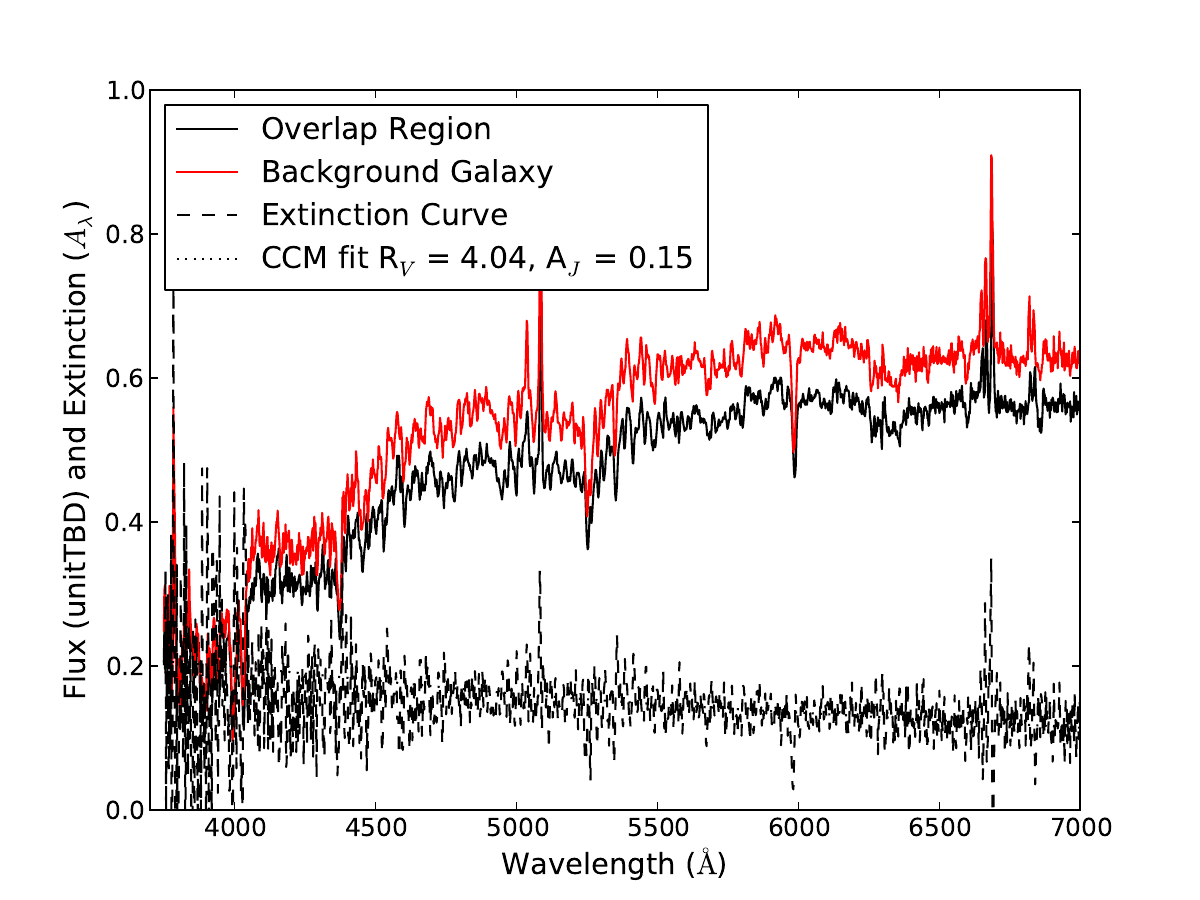}
\caption{A good example of two fiber spectra from the V500 cube: one fiber spectrum from the overlap region (OL, black solid line), 
its counterpart in a corresponding background fiber (BG, red solid line) and the resulting extinction curve (dashed line) and CCM fit to that extinction curve (dotted line).}
\label{f:cal:fib}
\end{center}
\end{figure}

\begin{figure*}
\begin{center}
\includegraphics[width=0.49\textwidth]{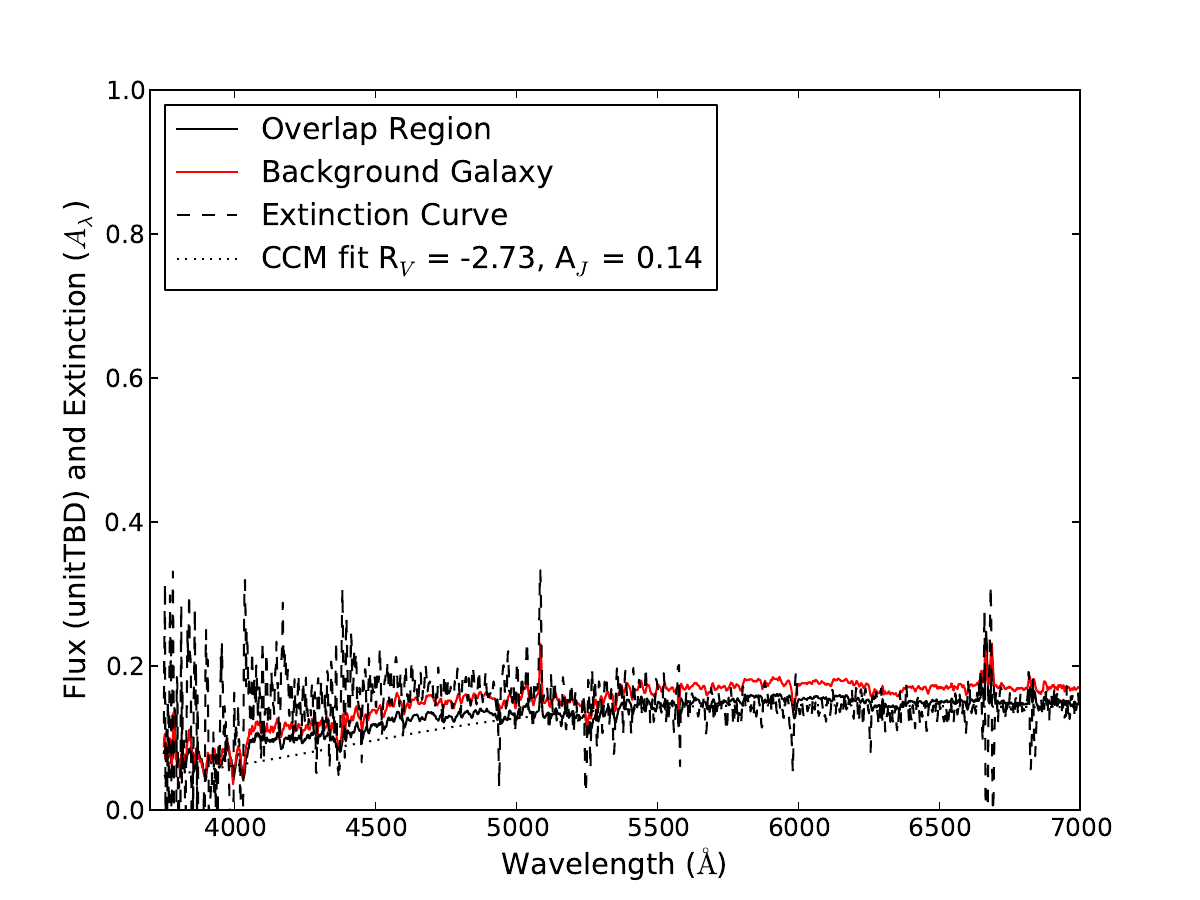}
\includegraphics[width=0.49\textwidth]{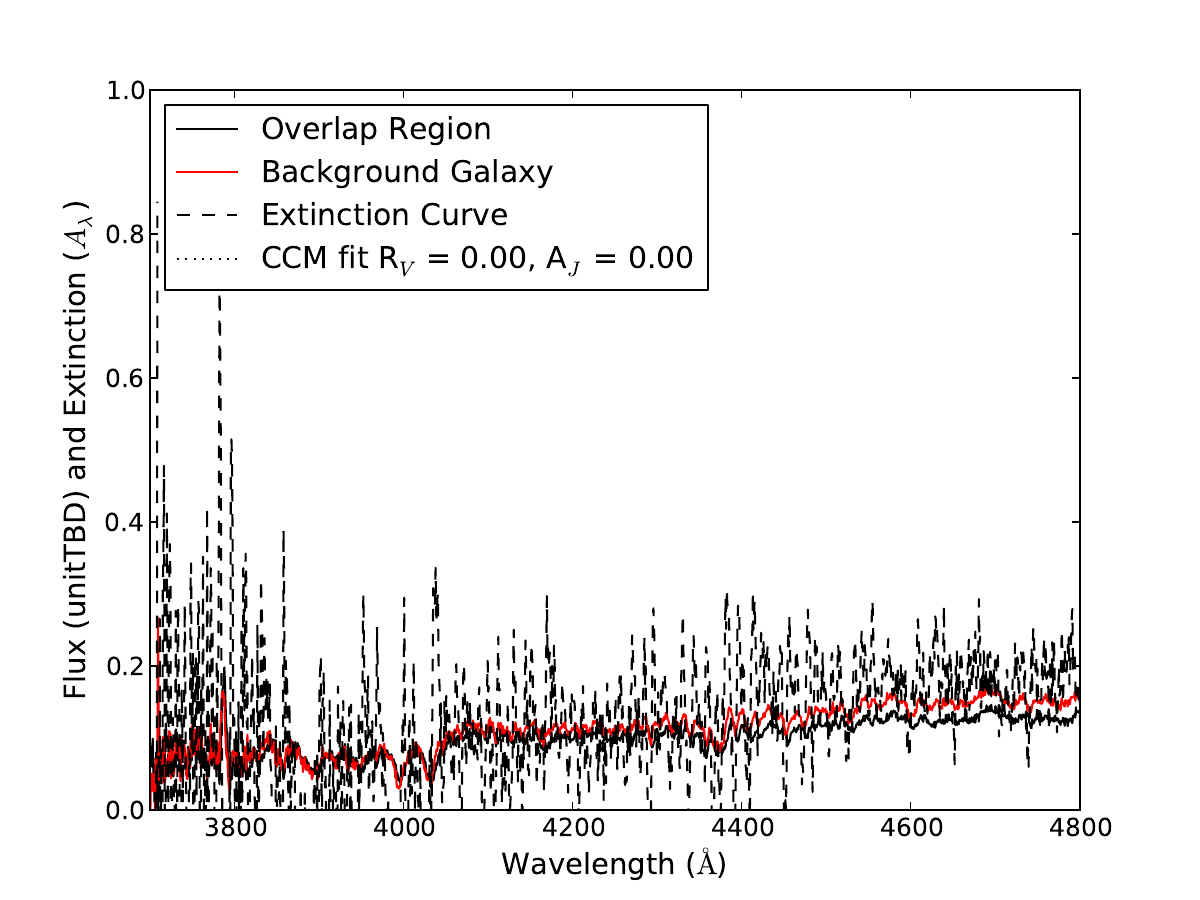}
\caption{The mean extinction curve of the overlap region (green box in Figure \ref{f:cal:fov}) in the V500 (left) and V1200 (right) cubes. 
This is to illustrate the effect of spatially smoothing an extinction signal over a spiral disc: little or any extinction is detected and 
no  slope in the extinction
}
\label{f:cal:meanspec}
\end{center}
\end{figure*}

\subsection{CALIFA data}
\label{s:cal:analysis}

The CALIFA data cubes were analysed fiber-by-fiber, assuming again that the light contribution from the foreground galaxy is negligible compared to the bright background galaxy (valid mostly in the outer disc). 
The HST results show where the approximation of no foreground contamination is applicable: where the outer disc of UGC 3995B projected against inner bulge of UGC 3995A.
However, we are somewhat forced into this approach, as the corresponding aperture is not available in the PPAK field-of-view (Figure \ref{f:cal:fov}). We limit the analysis to the same aperture as used for the HST data (green box in Figures \ref{f:hst:fov}) for ease of comparison. 

To obtain the extinction curve in each fiber in the aperture, we find the corresponding fiber in the background galaxy by rotating the position $180^\circ$ around the central position of the UGC 3995B (ra=116.03393, dec=29.247411). The extinction curve, $A(\lambda)$, follows from the ratio of the overlap spectrum, $OL(\lambda)$, in the aperture and the background galaxy spectrum, $BG(\lambda)$, in the corresponding fiber:
\begin{equation}
A(\lambda) = -1.086 \times ln \left({OL(\lambda) \over BG(\lambda)}\right)
\end{equation}

An example of the OL and BG fiber spectra and the resulting extinction curve is shown in Figure \ref{f:cal:fib}. 
The extinction curve in many individual fibers is of sufficient signal-to-noise to fit the extinction law parameterization of \cite{CCM} to each:
\begin{equation}
A(\lambda) = a(\lambda) A_V + b(\lambda) {A_V \over R_V}
\end{equation}
\noindent where $a(\lambda)$ and $b(\lambda)$ are the polynomials determined by \cite{CCM} and $A_V$ and $R_V$ are the normalization and slope of the fit. We compute the polynomials for the wavelength range and subsequently perform a linear fit to $A(\lambda)$ with $A_V$ and $R_V$ as free parameters with {\sc numpy.polyfit} in the {\sc python} environment.
We fit both CALIFA cubes over most of their respective wavelength ranges: we fit over the sub-range 5200--7300 \AA\ for the V500 low-resolution cube and 4000--4700 \AA\ for the high-resolution V1200 cube, to avoid edge-effects in the extinction slopes.

Figure \ref{f:cal:ccm} shows the overlap aperture and the values of the $A_V$ and $R_V$ for each fiber fit in both the low-resolution cube (V500) and high-resolution cube (V1200). 
The two normalization ($A_V$) maps agree qualitatively very well, although there are some patches of higher extinction in the Northern structure in the high-resolution and bluer cube (V1200). The higher values of the extinction slope normalization are, in part, due to a fit to a wavelength range blueward of the reference Johnson V band; the value is extrapolated for comparison, from the spectral range observed to Johnson V. (5500\AA). The lower-resolution V500 cube actually includes the Johnson V.
We noted in H09 that the foreground dust disc appears patchier in the bluer wavebands as well.
The result is robust with smoothing in the spectral direction (Figures \ref{f:cal:Avmap:r500} and \ref{f:cal:Avmap:r1200}).

In contrast, the $R_V$ maps are almost random and the results from both cubes compare poorly with each other. 
The random pattern of $R_V$ values persists in the fits of fiber spectra after these were spectrally smoothed to boost signal-to-noise (Figures \ref{f:cal:Rvmap:r500} and \ref{f:cal:Rvmap:r1200}), that is, these spectral data are not sufficient in sampling and S/N to trace the R parameter.

The distributions of $A_V$ and $R_V$ derived for both cubes are shown in Figures \ref{f:cal:Avhist} and \ref{f:cal:Rvhist}. The normalization behaves approximately as one can expect. The mean $A_V$ is zero, implying little to no sign of an extinction law in the overlap region.


Figure \ref{f:cal:meanspec} shows the mean extinction curves taken over the aperture for the V500 and V1200 cubes. In this case, we do not average in the spectral direction but spatially over the whole overlap aperture. The resulting curve is extremely flat as it combines known high and low extinction areas. It serves to illustrate how, if one takes a large aperture over a spiral disc, the extinction law always becomes essentially flat ($R_V=1$).

\subsection{Comparing HST and PPAK results}
\label{s:analysis:both}

The normalization maps follow the extinction map in Figure \ref{f:hst:Avmap} reasonably well: there are two spiral structures visible in the North and East of the aperture (complexes A \& C in Figure \ref{f:hst:Avmap}) but certainly not all the (spiral arm) structure visible in Figure \ref{f:hst:Avmap} can be identified in Figure \ref{f:cal:ccm}. The agreement between the results from both CALIFA cubes is good as well. 

The distributions of $A_V$ values in Figure \ref{f:cal:Avhist} for both CALIFA cubes show a different range of values than Figure \ref{f:hst:Avhist}; there are more negative values as well as a more gradual decrease. The V500 values peak at $A_V=-0.2$, the V1200 cube's distribution is closer to Figure \ref{f:hst:Avhist}; values extend closer to $A_V=2$ and there is a peak at $A_V=0.5$ but there is another peak at $A_V=0$. corresponding to the part of the aperture in Figure \ref{f:cal:fov} not covered by UGC 3995B. 

There are three differences between the two data-sets: spatial resolution, signal-to-noise in each resolution element, and the spectral range.
The spatial resolution of the HST data is significantly better than that of the CALIFA cube, even if the signal-to-noise is substantially lower as it is only broad-band information. Spectral coverage of the V500 curve includes the Johnson V filter, while the high-resolution V1200 cube does not.



\section{Discussion}
\label{s:disc}

UGC 3995 is only the second galaxy pair for which serendipitously, both HST and IFU data are available. However, unlike the pair analysed in \cite{Holwerda13}, we are certain this one is interacting strongly. The UGC 3995 pair has several advantages though: it is a factor four closer (z=0.015812 to z=0.06) than our other pair and the foreground galaxy is close to face-on. The spatial sampling of the foreground disc is therefore much better in UGC 3995: $\sim30$ pc for the HST (PSF $= 0\farcs1$) data. 
However, the CALIFA spatial resolution is $\sim3\farcs$ \cite[][]{Husemann13}, which translates to $\sim0.9$ kpc.

The physical scale of the CALIFA cubes is still at least a factor ten above the maximum we determined for recovering a Milky-Way Extinction Law from Hubble color-imaging of a few galaxy pairs \citep[$\sim$100 pc.][H09]{kw01a}\footnote{Unfortunately, no multifilter HST imaging is available for UGC 3995, or a similar extinction law measurement could have been made.}. If the slope of the extinction law is determined over both highly opaque and mostly transparent sections, the extinction law effectively becomes grey \citep[low values of $R_V$, see also][]{Natta84, kw01a, Holwerda09, Grosbol12}. 
Thus, the lower range in extinction curve slopes visible in Figure \ref{f:cal:ccm} can be explained by the mixing of transparent and opaque regions. 
In their comparison between galaxy-wide SEDs  in \cite{Wild11} find extinction laws very similar to the Milky Way one. The measured dust reddening in this case is dominated by the optically thin parts of the galaxy discs; the reddening of the observed light is accurately retrieved but the total light absorption is not (as opaque regions are effectively not included).

High values of $R_V$ may point to recently formed or reprocessed dust or just poor fits. If the high $R_V$ values are indeed real, they may point to cases where large grains have yet to coalesce or have been disintegrated by shocks or a strong ultraviolet radiation field (e.g., through shocks or recently formed stars).

In a comparison with the other galaxy pair we have both IFU (VIMOS) and HST data for \citep[][]{Holwerda09} and \cite{Holwerda13a}, two things stand out: the distribution of $A_V$ values and the range of $R_V$ values fit in the fibers. The spatial sampling by the VIMOS data is $\sim 0.9$ kpc (both seeing and sampling are $\sim$0\farcs7) so we expect a equally greyer relation: we find in a similar overlap $R_V$ values between 1 and 3 with sensible fits in most fibers. 
Here we find a much greater range of $R_V$ values. In part this maybe because the symmetry argument does not hold completely for UGC3445B anymore as it is in interacting. Asymmetry would compound any problematic fits to the extinction curves. Overlapping discs will have to be sampled more finely in the future if we are to recover the extinction curve's slope.

The distribution of extinction values in the HST data is in our view telling: it peaks not at the typical $A_V=0$ but $A_V=0.3$, even though a similar fraction of the foreground disc is overlapping and covered by HST and IFU observations.
 The redistribution is either the result of a strong asymmetry in UGC 3995A, the background galaxy, or the result of an actual redistribution of dust in the disc of UGC 3995B. The latter might take the form of moving dust from low to higher column densities or removing a large fraction of the diffuse dust disc (destruction of grains). Arguing in favor of the redistribution is the more extended distribution beyond $A_V = 0.4$ and the effectively empty inter-arm regions.

 
\section{Conclusions}
 \label{s:concl}
 
\begin{enumerate}
\item UGC 3995B shows several spiral extinction features (Figure \ref{f:hst:Avmap}).
\item The geometry of this system allows construction of an unusually complete transmission map from HST data at 6000 \AA, and from this radially-averaged
distributions of transmission for the whole disc and for interarm regions (Figure \ref{f:hst:T}).
\item The distribution of $A_V$ values in the HST extinction map peaks near $A_V = 0.3-0.4$, unlike the distribution observed in other galaxies (Figure \ref{f:hst:Avhist}).
\item Beyond this point, the distribution of $A_V$ values drops like an exponential: $N(A_V) = N_0 \times e^{(-A_V/0.5)}$. The 0.5 value is higher than typical for a spiral galaxy (Figure \ref{f:hst:Avhist}).
\item The radial distribution of $A_V$ values declines slowly (Figure \ref{f:hst:RAv}).
\item A map of the extinction constructed from PPEX IFU data-cubes shows the same spiral structure of the HST extinction map (Figures \ref{f:hst:Avmap} \& \ref{f:cal:ccm} and \ref{f:cal:Avmap:r500} \& \ref{f:cal:Avmap:r1200}).
\item The inferred extinction slope ($R_V$) maps do not display any structure and a range of values partly due to the sampling of the disc by fibers and possibly partly due to the reprocessing of dust grains in the interacting disc (Figures \ref{f:cal:ccm}, \ref{f:cal:Avhist} and \ref{f:cal:Rvhist}).
\end{enumerate}

Future applications of IFU observations could shed more light on the composition of late-type discs, provided the spatial sampling is sufficient. Ongoing efforts to equip existing telescopes with improved IFUs (e.g, the WEAVE instrument on WHT or the MUSE on VLT\footnote{See also \url{http://www.ing.iac.es/weave/moslinks.html}}) will have the spatial resolution, field-of-view, and spectral range to facilitate detailed mapping of the effective extinction law, and thus dust grain distribution, in occulting galaxy pairs.

\section*{Acknowledgements}

The authors would like to thank the anonymous referee for a thoughtful and comprehensive rapport. 
The lead author thanks the European Space Agency for the support of the Research Fellowship program.
This research has made use of the NASA/IPAC Extragalactic Database (NED) which is operated by the Jet Propulsion Laboratory, California Institute of Technology, under contract with the National Aeronautics and Space Administration. 
This research has made use of NASA's Astrophysics Data System. Plots were made using the {\sc matplotlib} environment in {\sc Python} \citep{matplotlib}.
This study makes uses of the data provided by the Calar Alto Legacy Integral Field Area (CALIFA) survey (\url{http://califa.caha.es/}). Based on observations collected at the Centro Astron\'omico Hispano Alem\'an (CAHA) at Calar Alto, operated jointly by the Max-Planck-Institut f\"ur Astronomie and the Instituto de Astrofisica de Andalucia (CSIC).

%

\clearpage
\newpage

\appendix

\begin{figure*}

\section{Smoothing the CALIFA cubes}

The following figures show the effect of smoothing the high (V1200) and low-resolution (V500) CALIFA cubes in the spectral direction
on the resulting CCM extinction curve fits, both the normalization ($A_V$) and slope ($R_V$).

\begin{center}

\includegraphics[width=0.32\textwidth]{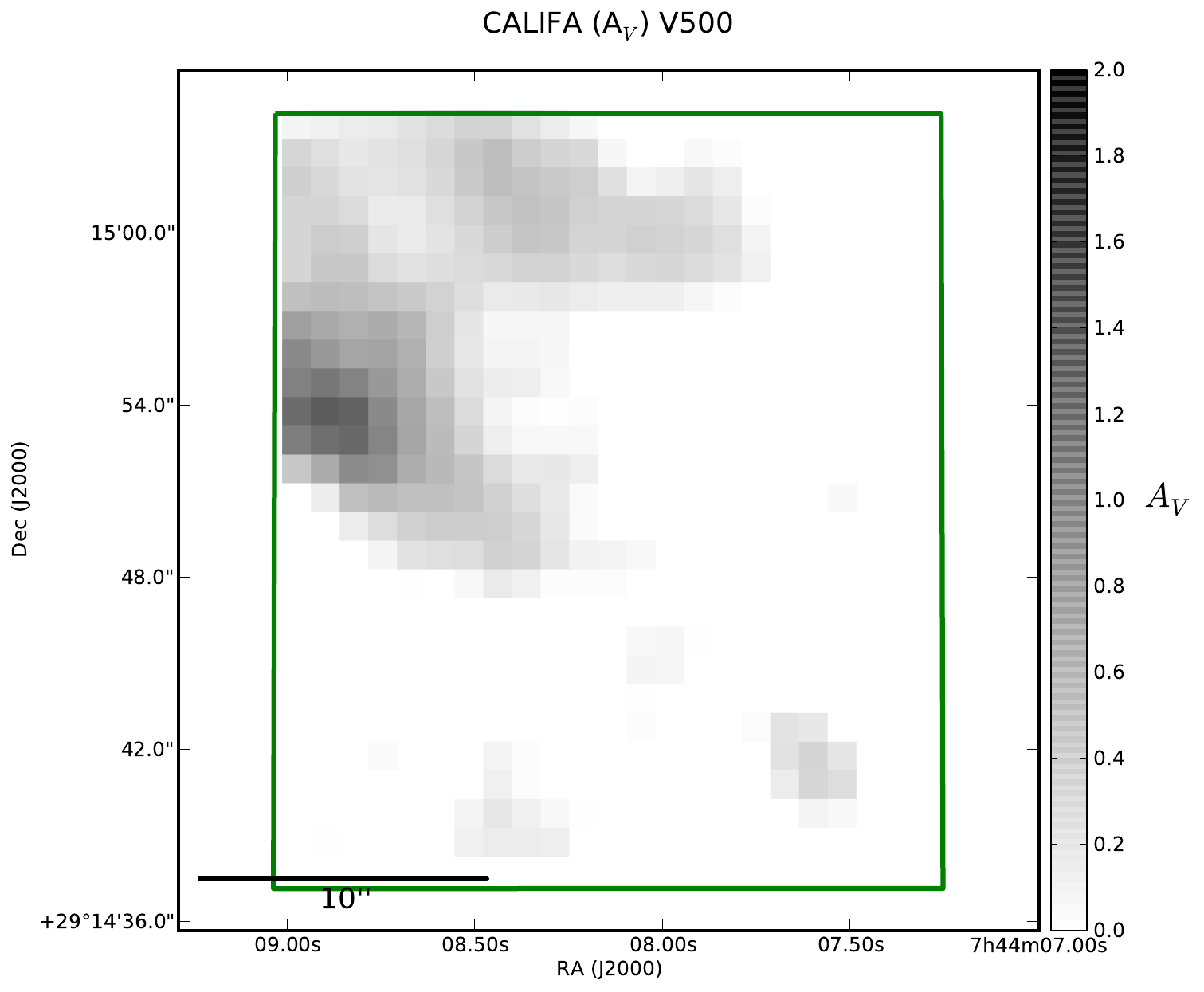}
\includegraphics[width=0.32\textwidth]{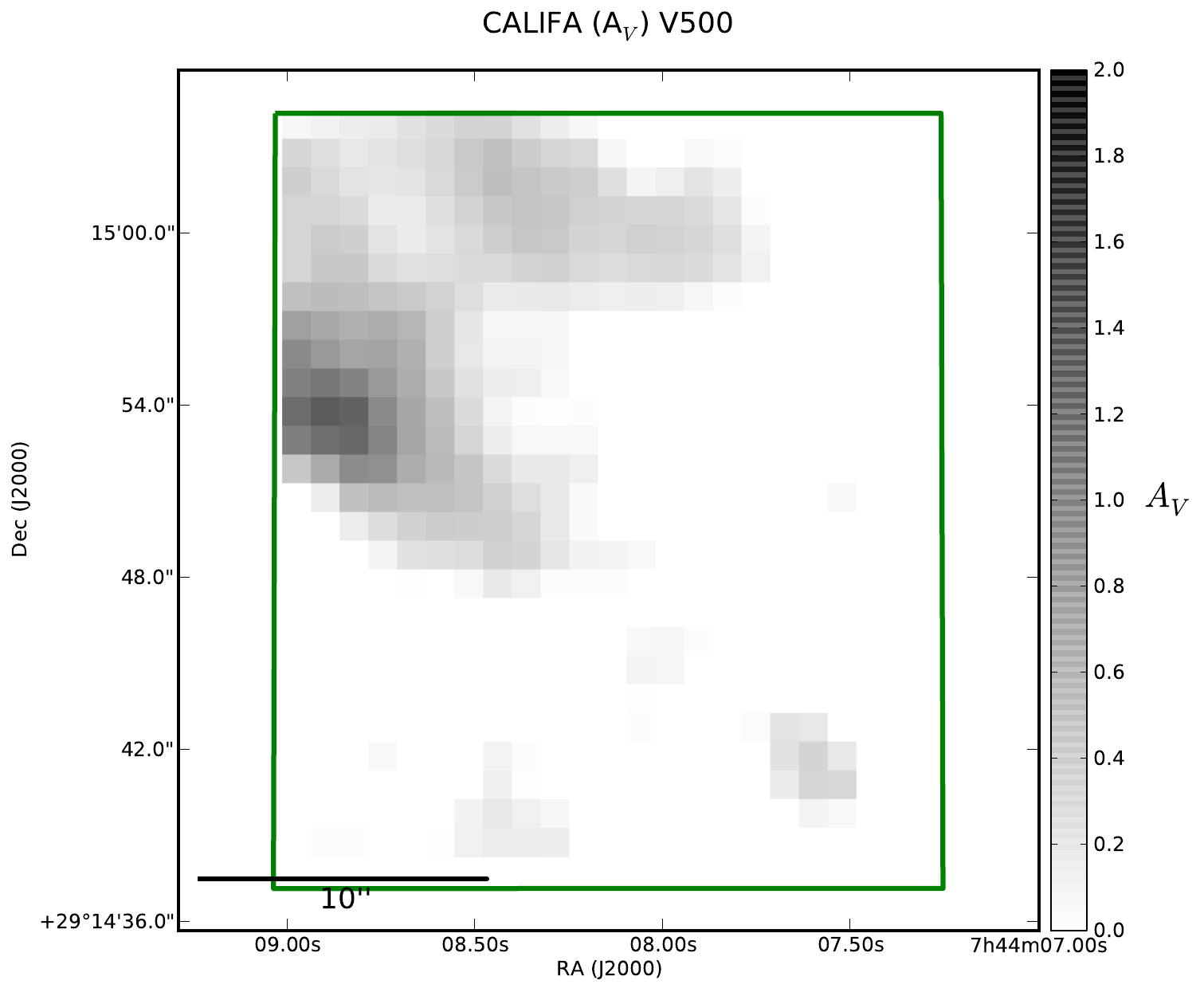}
\includegraphics[width=0.32\textwidth]{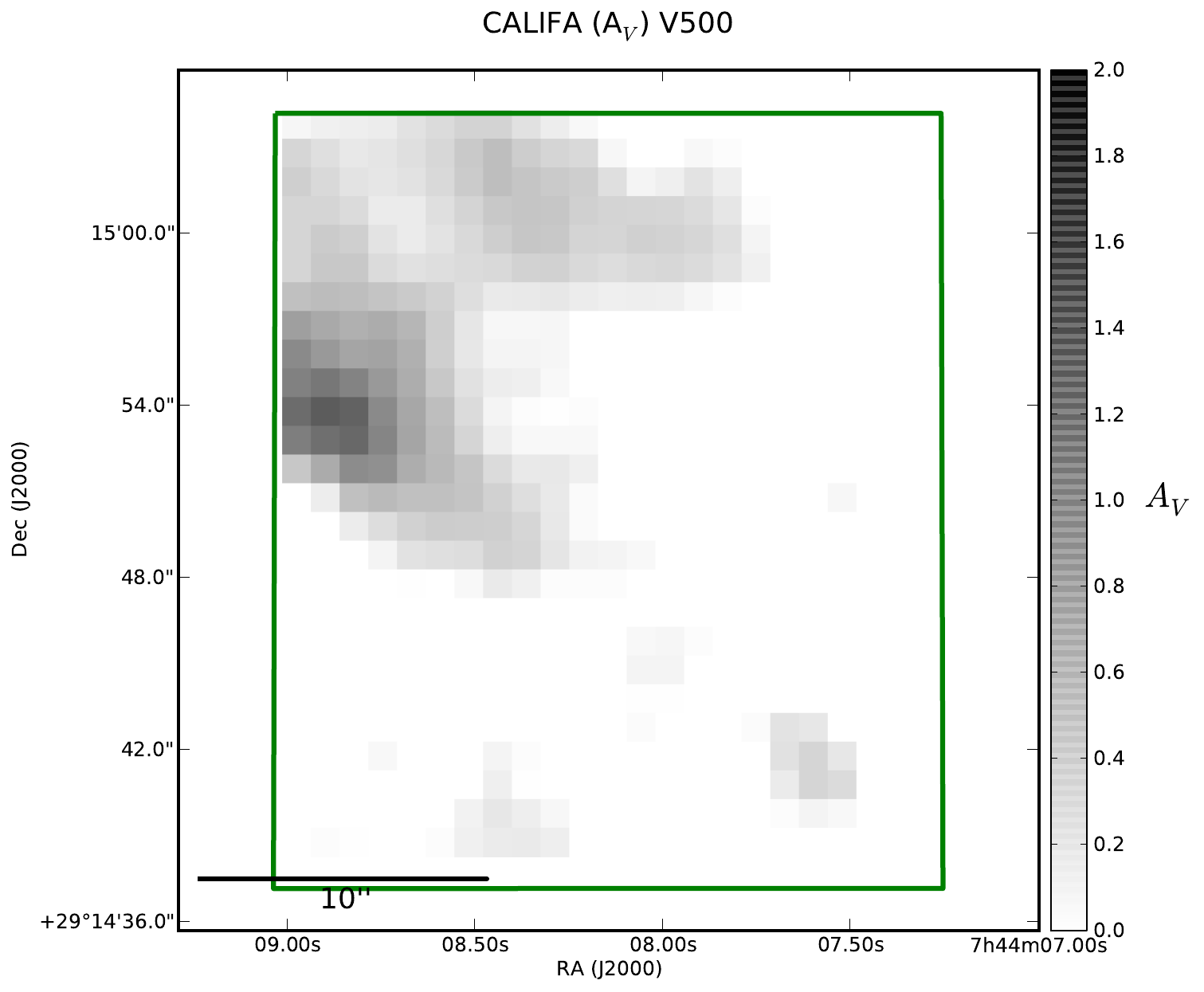}

\caption{The extinction in the overlap region (green box in Figure \ref{f:cal:fov}) in the V500 cube. The spectra were not smoothed  (left panel), and smoothed by 10 and 14 \AA\ (middle and right panels respectively) before a CCM extinction curve was fit to it. These panels show the normalization ($A_V$) of the fit. }
\label{f:cal:Avmap:r500}
\end{center}
\end{figure*}

\begin{figure*}
\begin{center}
\includegraphics[width=0.32\textwidth]{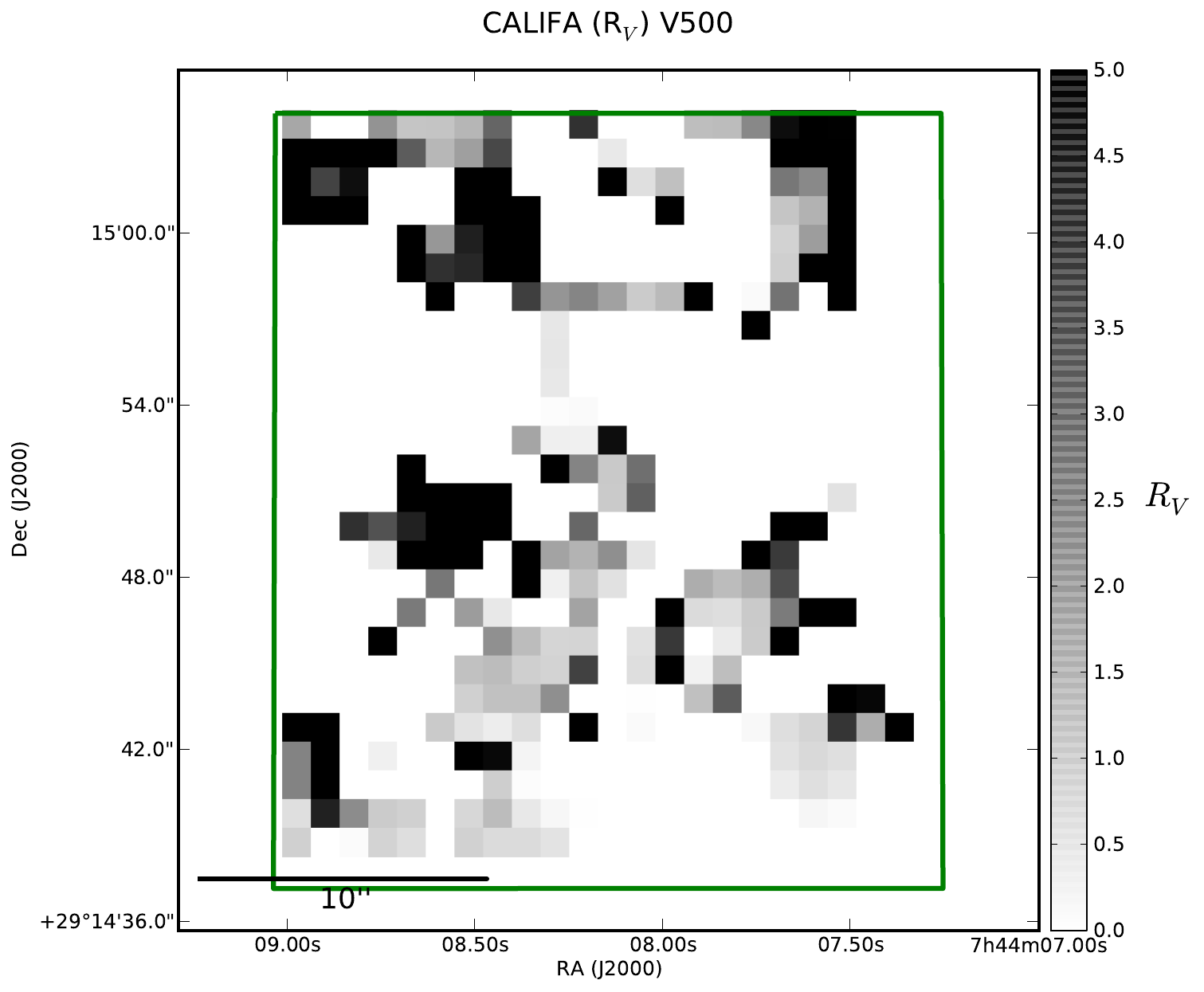}
\includegraphics[width=0.32\textwidth]{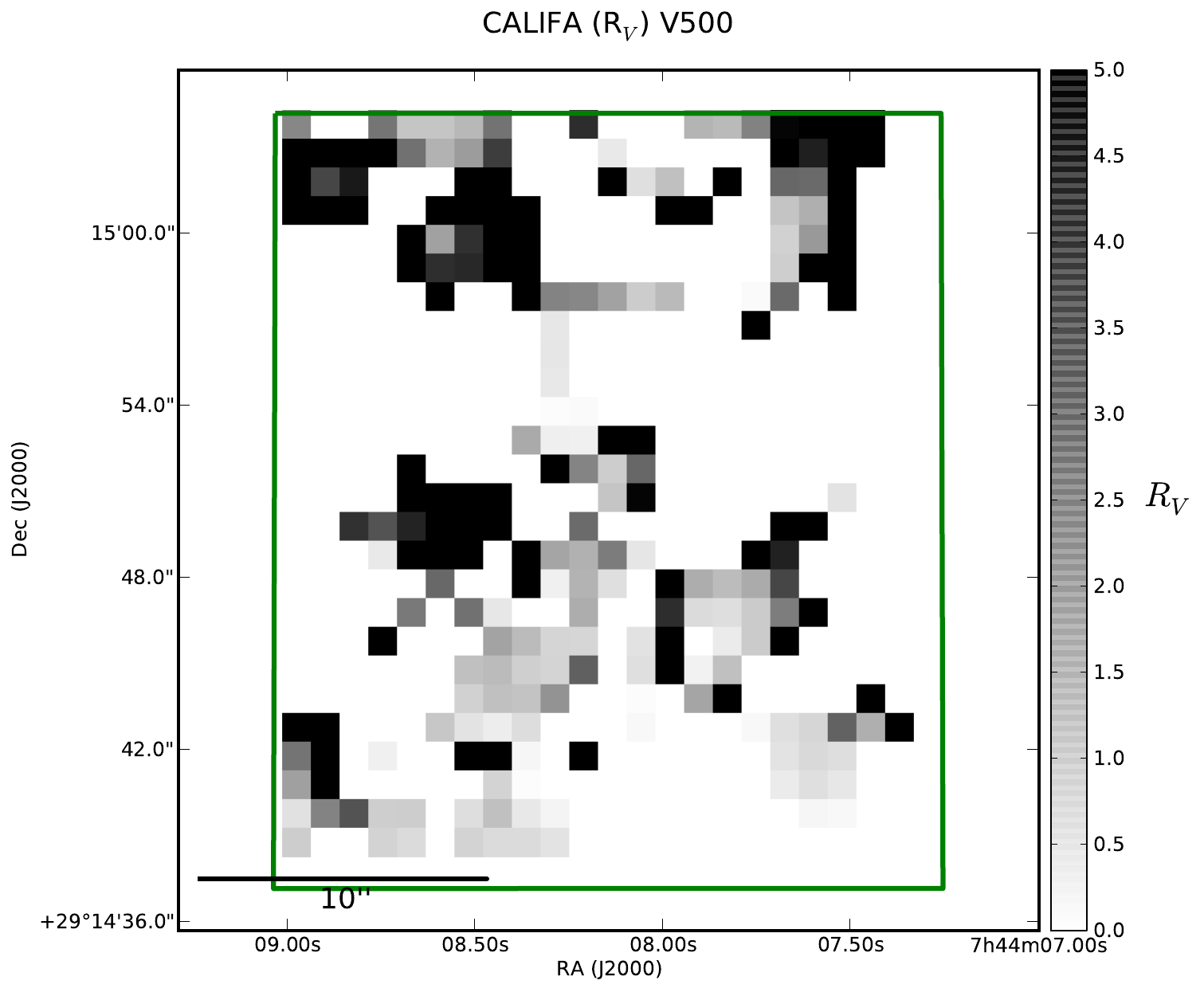}
\includegraphics[width=0.32\textwidth]{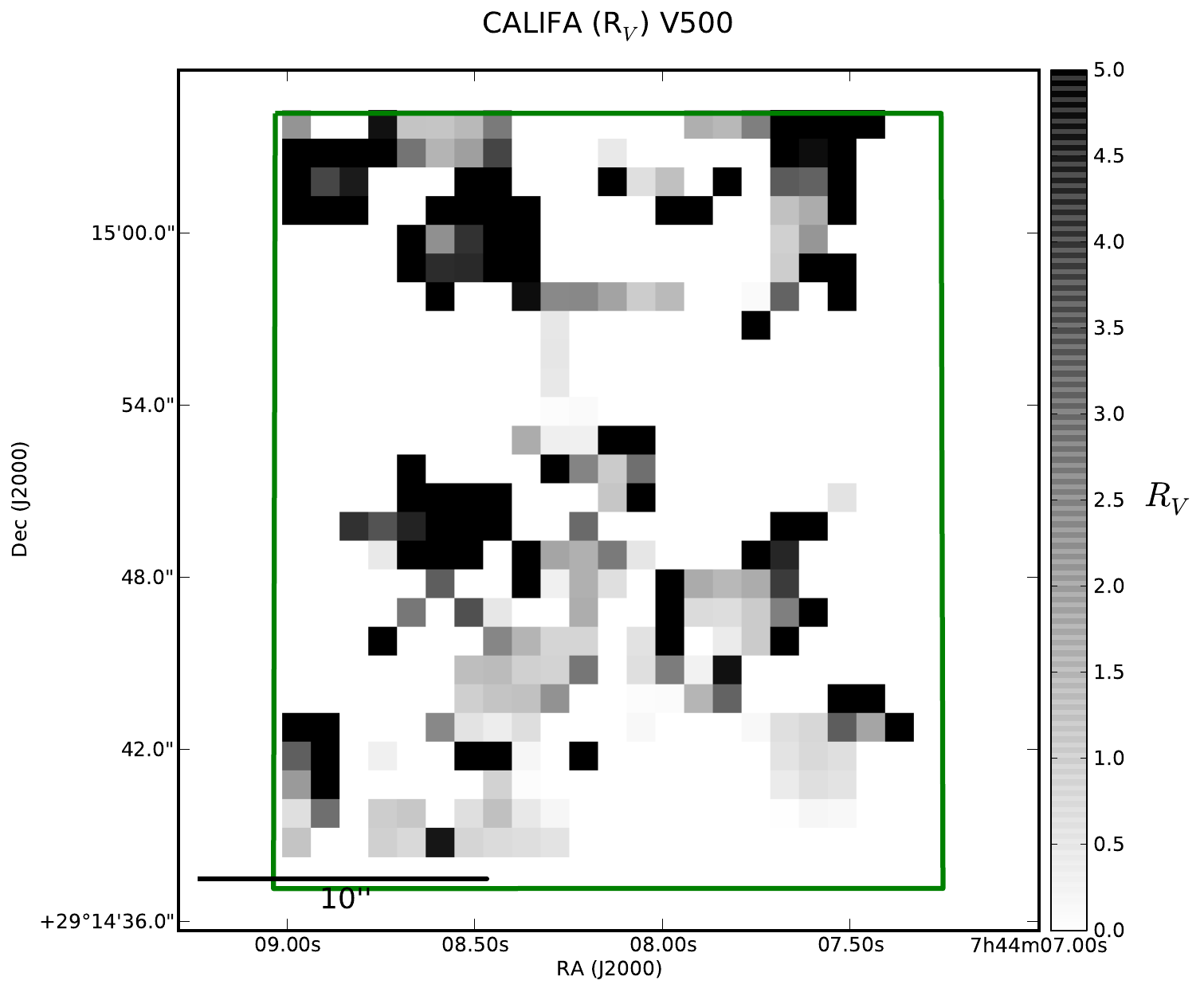}
\caption{The extinction law in the overlap region (green box in Figure \ref{f:cal:fov}) in the V500 cube. The spectra were not smoothed  (left panel), and smoothed by 10 and 14 \AA\ (middle and right panels respectively) before a CCM extinction curve was fit to it. These panels show the slope ($R_V$) of the fit.  }
\label{f:cal:Rvmap:r500}
\end{center}
\end{figure*}

\begin{figure*}
\begin{center}
\includegraphics[width=0.32\textwidth]{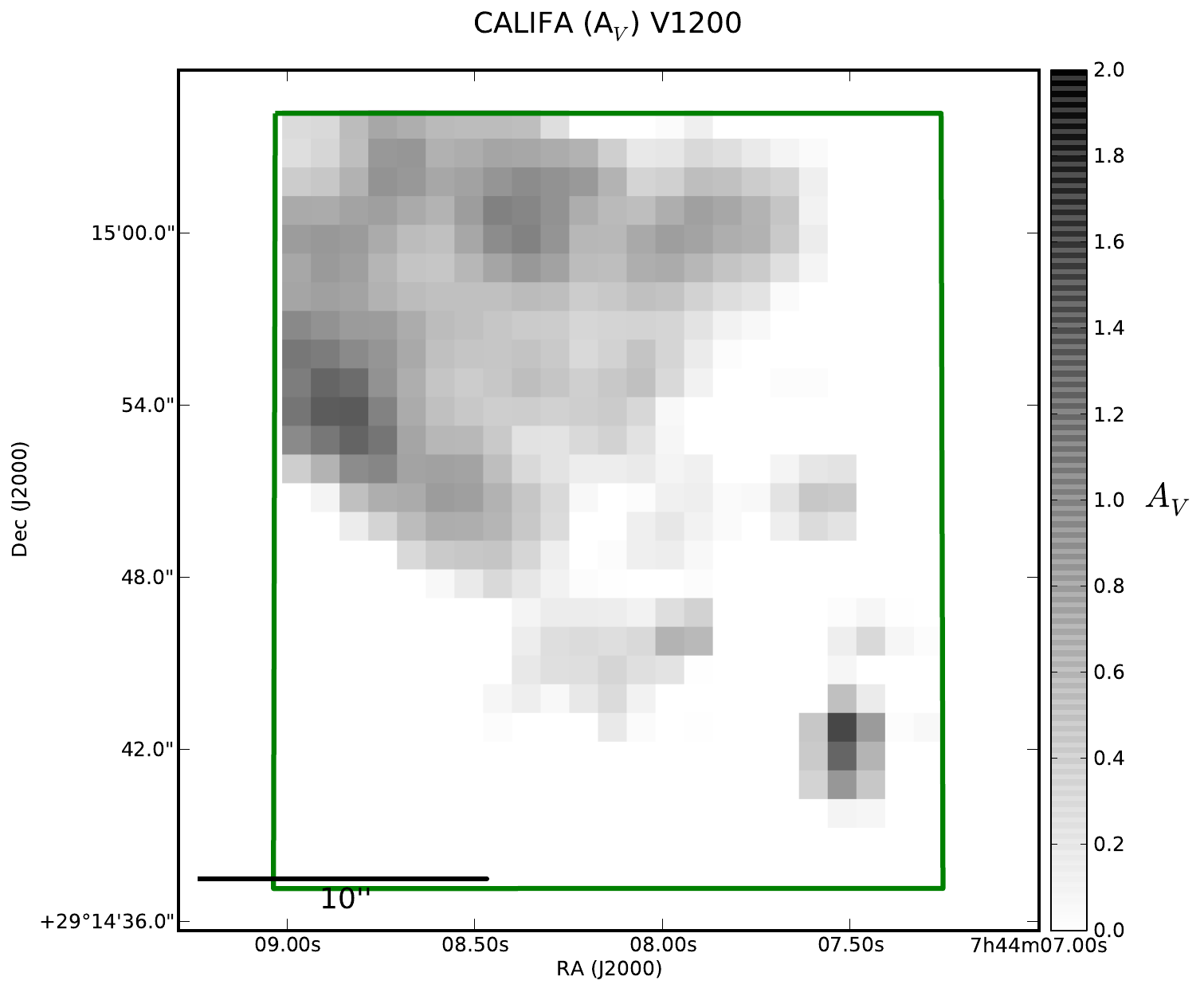}
\includegraphics[width=0.32\textwidth]{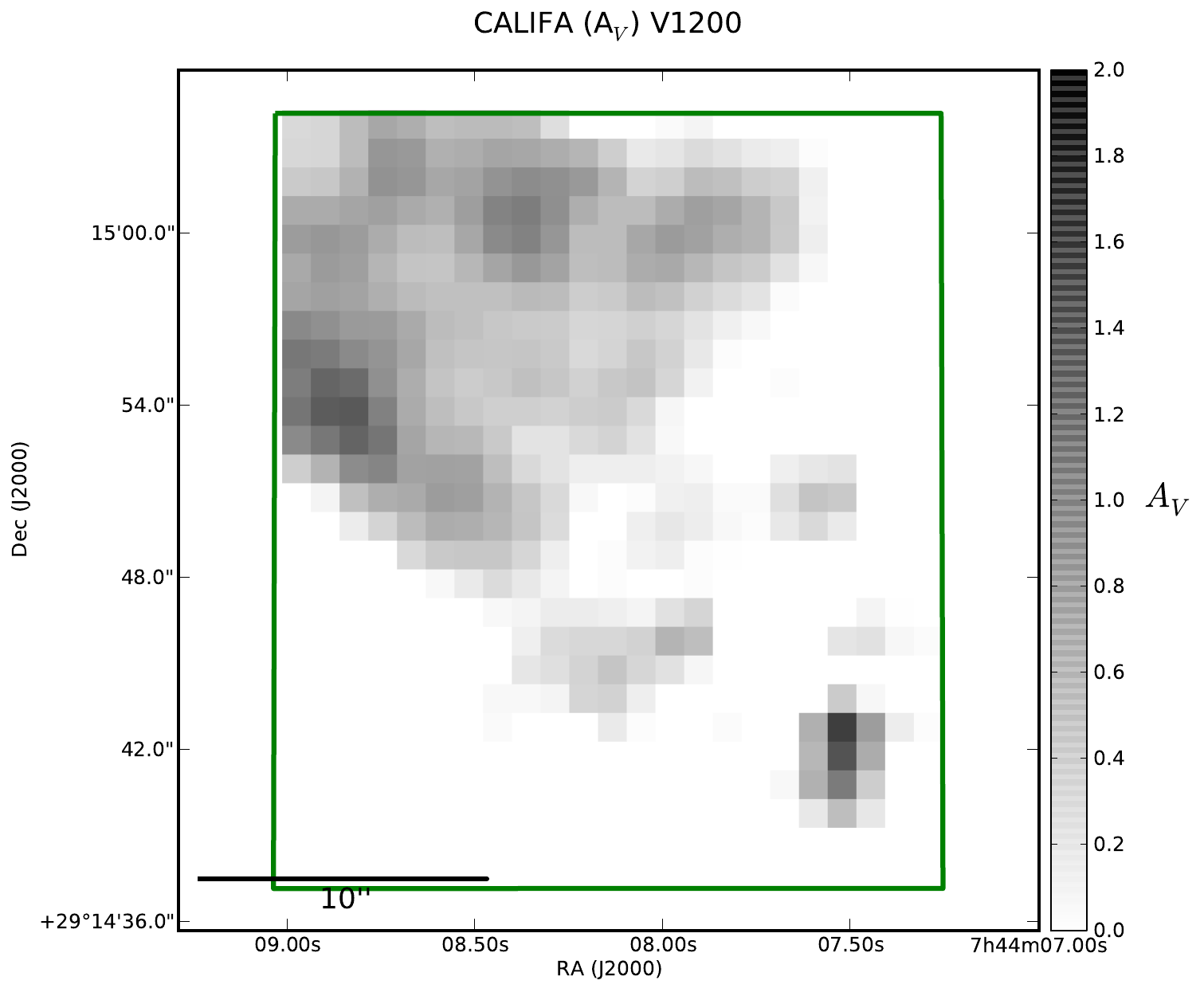}
\includegraphics[width=0.32\textwidth]{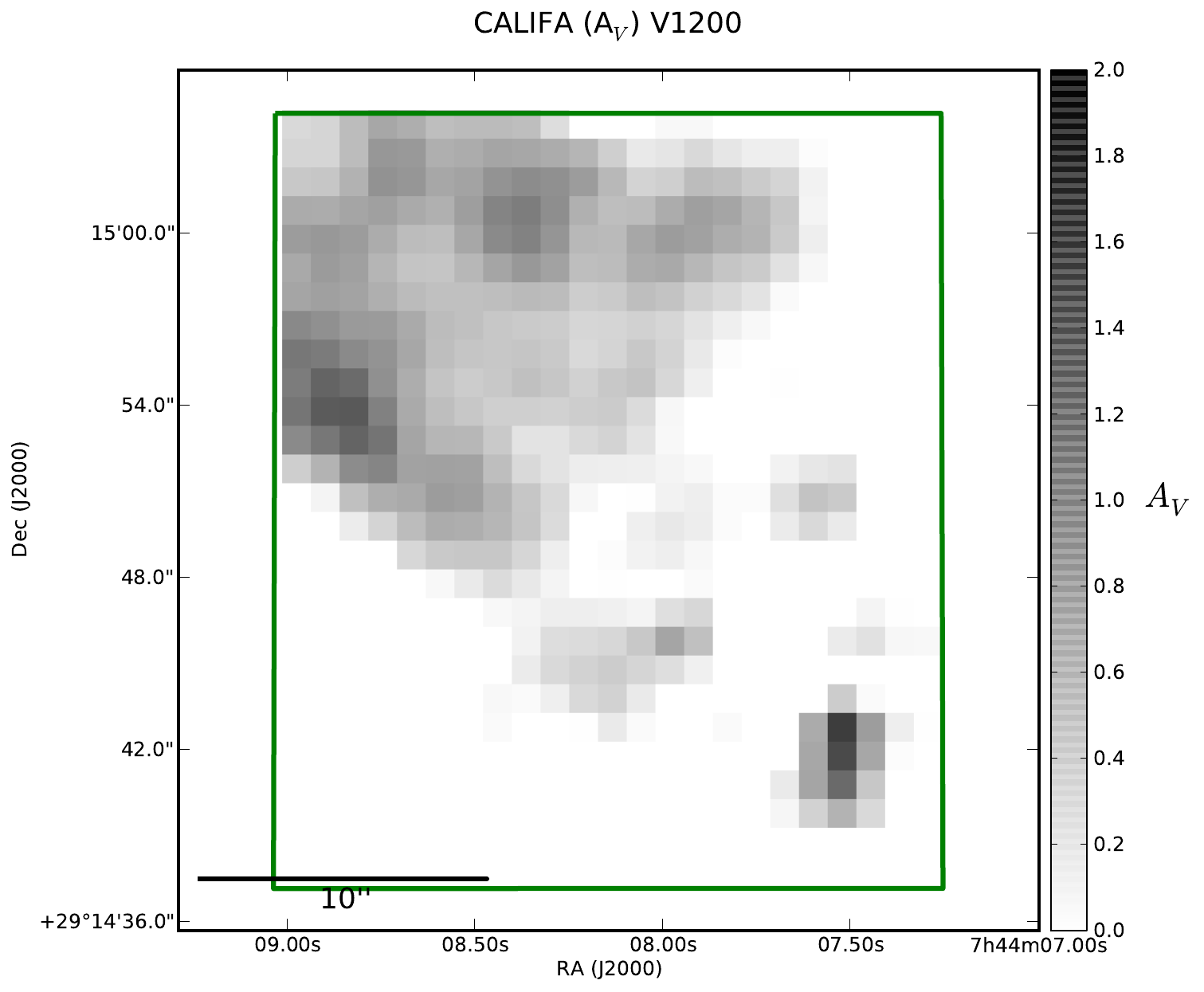}
\caption{The extinction in the overlap region (green box in Figure \ref{f:cal:fov}) in the first part of the V1200 cube.  The spectra were not smoothed  (left panel), and smoothed by 3.5 and 4.9 \AA\ (middle and right panels respectively) before a CCM extinction curve was fit to it.   These panels show the normalization ($A_V$) of the fit. }
\label{f:cal:Avmap:r1200}
\end{center}
\end{figure*}

\begin{figure*}
\begin{center}
\includegraphics[width=0.32\textwidth]{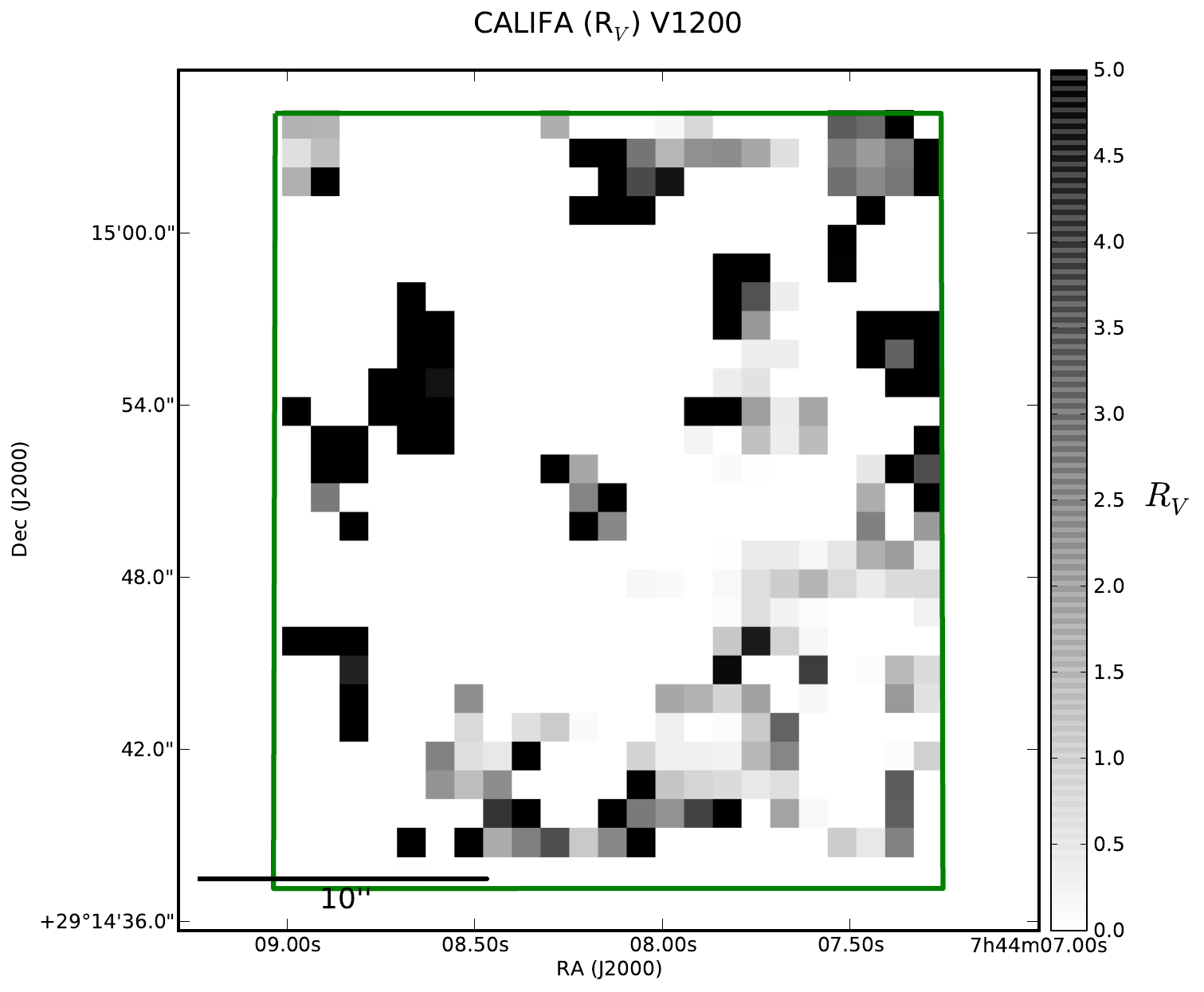}
\includegraphics[width=0.32\textwidth]{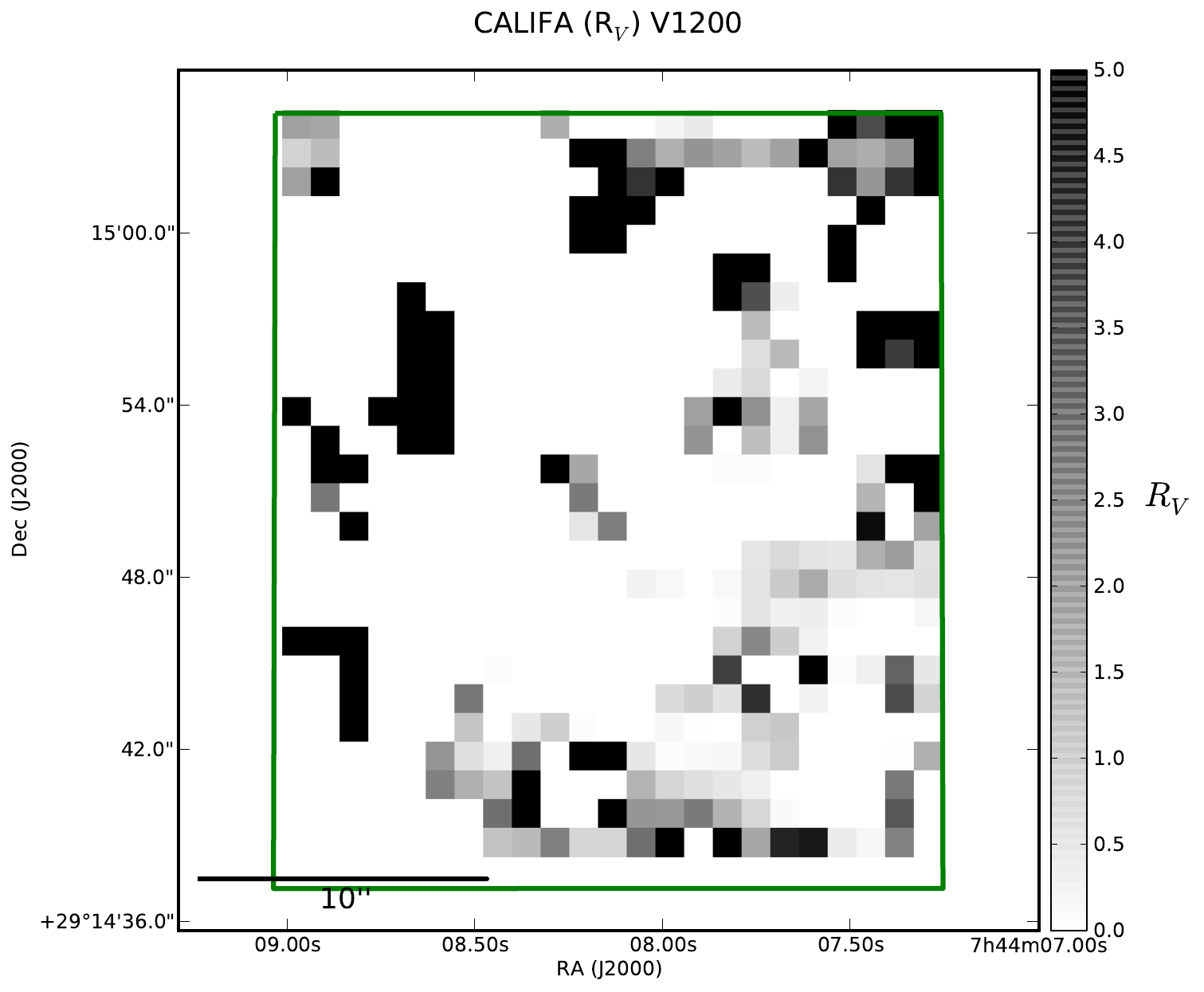}
\includegraphics[width=0.32\textwidth]{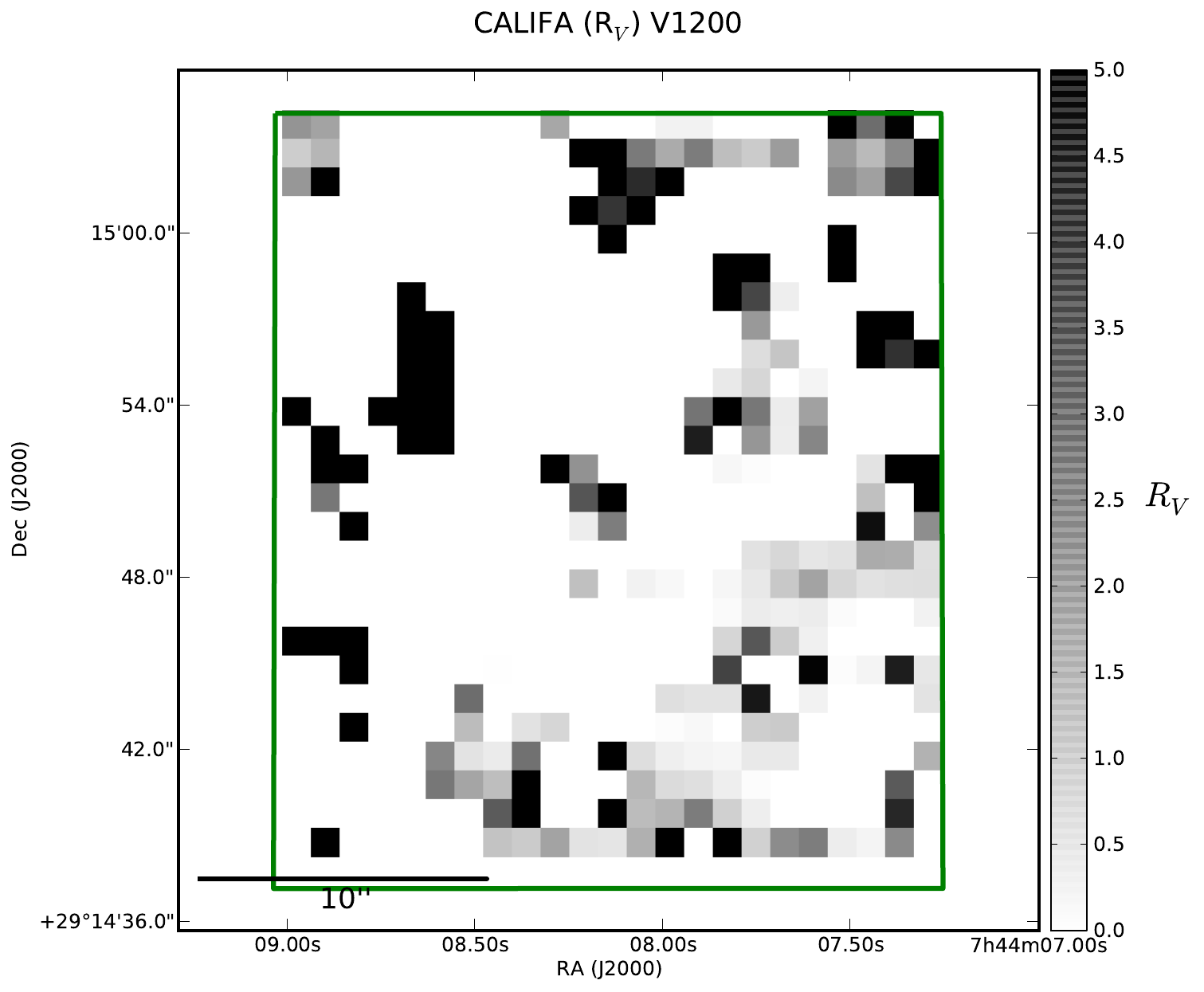}
\caption{The extinction curve in the overlap region (green box in Figure \ref{f:cal:fov}) in the first part of the V1200 cube.  The spectra were not smoothed  (left panel), and smoothed by 3.5 and 4.9 \AA\ (middle and right panels respectively) before a CCM extinction curve was fit to it.  These panels show the slope ($R_V$) of the fit.   }
\label{f:cal:Rvmap:r1200}
\end{center}
\end{figure*}

\end{document}